\documentclass[pmlr]{jmlr}

\usepackage{longtable}

\usepackage{booktabs}
\usepackage[load-configurations=version-1]{siunitx} 
\usepackage{floatrow}

\usepackage{mathtools}
\usepackage{pgfplots}
\usepackage{color}
\usepackage{tabu}
\usepackage{tikz}
\usetikzlibrary{backgrounds}
\usetikzlibrary{bayesnet}

\newcommand{\predictors}{{\mathbf{P}}}
\newcommand{\scores}{{\mathbf{S}}}

\newcommand{\makeanonymous}[1]{{[ hidden ]}}

\newcommand{\miss}{{\color{red} $\times$}}

\newcommand{\chedit}[1]{{\color{black} #1}} 

\theorembodyfont{\upshape}
\theoremheaderfont{\scshape}
\theorempostheader{:}
\theoremsep{\newline}

\title[Simultaneous Measurement Imputation and Outcome Prediction for ATR]{Simultaneous Measurement Imputation and Outcome Prediction for Achilles Tendon Rupture Rehabilitation}
\author{\Name{Charles Hamesse}\thanks{Both authors contributed equally to this manuscript.} \Email {charles.hamesse@mil.be} \vspace{2pt}\\
\addr 
\begin{tabu} to \textwidth {@{}X[l]  X[r]@{}}
KTH Royal Institute of Technology  & Royal Military Academy \\
Stockholm, Sweden  & Brussels, Belgium 
\end{tabu}
\AND
\Name{Ruibo Tu}{\color{blue}\footnotemark[1]} \Email {ruibo@kth.se} \\
\addr 
KTH Royal Institute of Technology\\
Stockholm, Sweden 
\AND
\Name{Paul Ackermann} \Email {paul.ackermann@sll.se} \\
\addr 
Karolinska University Hospital\\
Stockholm, Sweden 
\AND
\Name{Hedvig Kjellstr\"om} \Email{hedvig@kth.se} \\
\addr 
KTH Royal Institute of Technology\\
Stockholm, Sweden 
      \AND
\Name{Cheng Zhang} \Email{Cheng.Zhang@microsoft.com} \\
\addr 
Microsoft Research\\
Cambridge, UK 
} 

\begin{document}

\maketitle

\begin{abstract}
Achilles Tendon Rupture (ATR) is one of the typical soft tissue injuries. Rehabilitation after such a musculoskeletal injury remains a prolonged process with a very variable outcome. Accurately predicting rehabilitation outcome is crucial for treatment decision support. However, it is challenging to train an automatic method for predicting the ATR rehabilitation outcome from treatment data, due to a massive amount of missing entries in the data recorded from ATR patients, as well as complex nonlinear relations between measurements and outcomes. In this work, we design an end-to-end probabilistic framework to impute missing data entries and predict rehabilitation outcomes simultaneously. We evaluate our model on a real-life ATR clinical cohort, comparing with various baselines. The proposed method demonstrates its clear superiority over traditional methods which typically perform imputation and prediction in two separate stages.
\end{abstract}

\section{Introduction}
Soft tissue injuries, such as Achilles Tendon Rupture (ATR), are increasing in recent decades \citep{Huttunen2014AcuteAT}. Such injuries require lengthy healing processes with abundant complications, which can cause severe incapacity in individuals. Influences of various factors such as  patient demographics and different treatment methods are not clear for the rehabilitation outcome due to large variations in symptoms and the long healing process. Additionally, many medical examinations are not carried out for a large portion of patients since they can be costly and/or painful. Thus, accurately predicting the ATR rehabilitation outcome at different stages using existing measurements is highly interesting, and can be used for decision support for practitioners. Moreover, ATR is one example of a wider class of medical conditions. In these situations, patients first need acute treatments, then go through a long-term and uncertain rehabilitation process \citep{deficits10yrhorstmann}. Decision support tools for practitioners are in general of great need, and outcome prediction plays an important role in the medical decision-making. 

Predicting ATR rehabilitation outcomes is extremely challenging for both medical experts and machines. This is mainly due to large number of noisy or absent measurements from various medical instruments. Medical tests and outcome scores for ATR involve a large variety of metrics. The total number of those metrics is on the magnitude of hundreds. However, only a subset of all possible medical measurements are used for a patient. Thus, the observations are very sparse. 
In this work, we use an ATR cohort which is collected from multiple hospitals in the past five years. 
The sparsity of this cohort is the consequence of several phenomena: firstly, they are aggregated from different studies realized by different clinicians who have different procedures; 
secondly, some measurements can be painful, costly or time-consuming such that not all patients are willing to take them. Such phenomena are common in many medical cohorts. Moreover, among those measurements, many are noisy and establish highly non-linear relationship to the rehabilitation outcome, which makes the outcome prediction task difficult.

Leveraging data-driven approaches, we design machine learning models to predict potential ATR rehabilitation outcomes with sparse and noisy data for patients, and provide decision support for practitioners.
In particular, we develop a probabilistic model to address two problems at once: imputing the missing values of the costly medical measurements for patients, and predicting patients' final rehabilitation outcome. We focus on predicting the ATR rehabilitation outcome, while our framework can be further applied to a wider domain of conditions beyond ATR.

\paragraph{Technical significance.}
We propose a novel probabilistic framework where probabilistic matrix factorization is combined with a Bayesian Neural Network (BNN) for rehabilitation outcome prediction with the noisy and sparse dataset. Our method shows clear improvement for this task comparing to traditional methods. 
For outcome prediction with such a cohort, traditional methods commonly need two stages: Firstly, missing values are imputed using methods such as mean-imputing or zero-imputing; secondly, a linear model is used to predict the outcome with the imputed data \citep{arverud2016ageing, bostick2010factors}. These methods commonly lead to a low prediction quality because of the low imputation quality and the linear relationship assumption between measurements and outcomes.

Our framework simultaneously imputes the missing values and predicts the rehabilitation outcomes.  We first adopt a probabilistic latent variable model to predict the missing entries in the hospital stay measurements. Prediction based methods, such as using latent variable models, in general, demonstrate superior performance for data imputation compared to traditional methods such as mean imputation \citep{scheffer2002dealing,buuren2010mice,keshavan2010matrix, ma2018eddi}. These latent variables summarize patients' underlying health situation in a low-dimensional space and impute the missing entries based on the patient status. 
We then combine the latent variable imputation model with BNN to predict ATR rehabilitation outcomes as one integrated probabilistic model. BNN is highly flexible, thus can handle non-linear relationships between rehabilitation outcomes and measurements. Moreover, our model is fully probabilistic, thus provides uncertainty estimation of the prediction results. In an end-to-end manner, our framework provides significant improvement in the clinical standard. 

\paragraph{Clinical relevance.}
ATR rehabilitation is a prolonged process with unpredictable variation in the individual long-term outcome. The optimal and individualized rehabilitation protocol is unknown and therefore inappropriate treatments may often be provided leading to worse outcome for the patient and increased cost for society. In this case, prediction of ATR rehabilitation outcome can shorten the healing process by helping clinicians to choose effective treatments based on varying patient characteristics. 

There is a large number of ATR treatments and rehabilitation protocols and also assessments made on the patients. Choosing suitable treatments for patients is still challenging. Moreover, different measurements may vary in price and time to perform. Given imputed values for measurements and the predicted values for outcomes with calibrated uncertainty using our method, the clinicians can make decisions on the patient treatment and monitoring more easily. For example, if the model predicts an unobserved measurement value with high confidence and the predicted value is in a clinical normal range, the clinician does not need to measure this value anymore. Thus, time and cost are saved by not performing unnecessary medical assessments. Otherwise, if the prediction indicates any abnormal situation or high uncertainty for an important measurement, it is worthwhile to apply this medical instrument and obtain the measurement value for this patient. The predicted outcome also helps the clinician to better estimate the patient status in general and aids the treatment decisions.

The paper is structured as follows: We discuss related work (Section \ref{sec:related}) and describe the ATR cohort (Section \ref{sec:cohort}). We then introduce the proposed model (Section \ref{sec:methods}). Finally we evaluate our proposed method against multiple baselines. The experimental results demonstrate clear improvement for ATR rehabilitation outcome prediction using our proposed model (Section \ref{sec:exp}).

\section{Related Work}
\label{sec:related}
Our work focuses on utilizing machine learning methods for ATR rehabilitation outcome prediction. We use a latent variable model based on probabilistic matrix factorization to address the missing entry problem, and then use the estimated patient state to predict the rehabilitation outcome through BNN. There is very limited work on using machine learning to address the ATR outcome prediction. We revisit the related work in the following three aspects: ATR analysis, AI in a generic health-care setting and missing value imputation, which is a key component for this type of applications. 

\paragraph{Achilles tendon rupture analysis.} Numerous studies have been carried out on understanding the treatment and rehabilitation of ATR due to its importance in health-care. However, most studies are performed with a clinical approach, and use traditional statistic analysis, typically linear regression. Machine learning based approaches have not been widely adopted in the field of ATR research. As such, tools for rehabilitation outcome prediction using machine learning are of great interest. Here, we briefly review some related work on ATR.

\citet{olsson2014ability} employ linear regression to predict the rehabilitation outcome using variables such as age, sex, body mass index (BMI) or physical activity. The result shows that using traditional statistical models such as linear regression yields a limited prediction ability despite having a wide range of clinically relevant variables. A more recent study shows that assessing clinical markers of tendon callus production (procollagen type I N-terminal propeptide (PINP) and type III N-terminal propeptide (PIIINP)) shortly after operation can help improve the prediction of long-term patient-reported outcomes applying multiple linear regression on Achilles Tendon Total Rupture Score (ATRS) one year post-injury \citep{Ackermann16}. Additionally, microcirculation in the tendon was also shown to be a strong predictor of the patient outcome after ATR \citep{praxitelous2017microcirculation}.  Although insightful, this research also shows that some accurate measurements, such as microcirculation, can be expensive or difficult to obtain. Therefore, utilizing a large range of cost-efficient data to predict the rehabilitation outcome is desirable.

\paragraph{AI in health-care.} There is a broad spectrum of machine learning methods used for generic medical applications. 
When dealing with large amounts of data, deep learning algorithms show promising results. For example, long short-term memory networks (LSTM) and convolutional neural networks (CNN) has been applied to various clinical tasks such as mortality prediction in ICU setting where the data are often gathered from sensor readings \citep{suresh2017clinical, chalapathy2016bidirectional, jo2017combining, purushotham2017benchmark}. 

However, health-care datasets often have a limited number of patients with large numbers of variables from different instruments. This often leads to datasets with many missing entries. At the same time, being able to encode existing medical research results in new models and providing interpretable results are desirable features in many health-care related applications. In this case, probabilistic models are needed. Depending on the medical context, different types of models are used. For example, \citet{lasko2014efficient} employs Gaussian processes to predict irregular and discrete medical events. \citet{framework-indiv} design a hierarchical latent variable model to predict the trajectory of an individual's disease. These models are developed for different medical contexts and are not directly applicable to our application setting. 
In this work, we use ATR as an example and design a model to predict patients' rehabilitation outcomes after acute treatments. 

\paragraph{Missing value imputation.} Most real-life medical cohorts have a large amount of missing values. Traditional methods such as zero imputation or mean imputation ease the analysis but may lead to low imputation accuracy. 
For the datasets with missing values, matrix factorization based methods are shown to be effective for many missing value imputation applications \citep{shi2016temporal,troyanskaya2001missing}, and frequently used for other applications of the matrix completion problem, i.e., collaborative filtering \citep{ocepek2015improving}. Many efficient algorithms have been proposed, such as Singular Value Thresholding (SVT) \citep{cai2010singular}, Fixed Point Continuation (FPC) \citep{ma2011fixed}, and Inexact Augmented Lagrange Multiplier (IALM) \citep{lin2010augmented}. Typically, these methods construct a matrix factorization objective and optimize it using traditional convex optimization techniques. Extensions, such as Singular Value Projection (SVP) \citep{jain2010guaranteed} and OptSpace \citep{keshavan2009gradient} consider observation noise in the objective. However, the sparse dataset can damage the performance of matrix factorization based methods \citep{mnih2008probabilistic}. In this case, probabilistic matrix factorization \citep{mnih2008probabilistic} is an alternative solution for sparse and imbalanced datasets. 

In this work, we combine a probabilistic matrix factorization approach similar to that of Matchbox \citep{matchbox}, with a supervised learning approach using models such as Bayesian neural networks \citep{neal2012bayesian}. Therefore, we can impute missing values based on latent patient traits with the sparse ATR dataset and predict the rehabilitation outcome in an end-to-end probabilistic framework.
\section{Cohort}
\label{sec:cohort}
In our work, the cohort is a real-life dataset  collected from multiple previous studies by an orthopedic group \citep{valkering2017functional,domeij2016ageing}. A snapshot of the dataset is shown in Figure \ref{dataset-snapshot}. There are $442$ patients in the dataset ($N =442$). The number of measurements is $M = 297$, and the number of the outcome scores is $S = 63$. We denote the first $N \times M$ part of the dataset as the \textit{predictors}, $\predictors$, and the second $N \times S$ part as the \textit{scores}, $\scores$. The percentage of missing values is 69.5\% in the predictors and 64.2\% in the scores. 

\begin{figure}[th]
{\footnotesize	
\begin{center}
\begin{tabular}{ c | c|  c | c | c | c | c| c| c| c | c  }
& Length & Weight & $\hdots$ & DVT\_2& $\hdots$ &ATRS\_12\_stiff\\
 \hline
1&  190& 79.8& $\hdots$& \miss& $\hdots$& 8\\
2&  \miss& 76.5& $\hdots$& 0& $\hdots$& \miss \\
3&  \miss& \miss& $\hdots$& 1& $\hdots$& 10 \\
4&  178&96.7& $\hdots$& 0& $\hdots$& \miss\\
\end{tabular}
\end{center}
}
\caption{A snapshot of the Achilles Tendon Rupture (ATR) cohort. 
Each row represents a patient's medical record and each column represents a measurement. 
In this example, DVT\_2 refers to the presence of deep venous thrombosis after two weeks and the ATRS\_12\_stiff measurements refer to the Achilles Tendon Rupture Score (ATRS) metrics of stiffness after 12 months. {\color{red}$\times$} indicates the entry is missing.}
\label{dataset-snapshot}
\end{figure}

\paragraph{Problem setting.} We review a typical case of patient journey first and then introduce the problems. ATR patients typically go to hospital to get a treatment immediately after an injury. There, their demographic data are registered. As part of the treatment process, they go through a number of tests from various medical instruments. Due to the complexity of these tests (e.g. in terms of time, cost, pain, invasiveness, accuracy), not all patients go through the same procedure. This leads to a lot of missing data and a lot of variation in which measurements are missing. After the treatment, patients are discharged from the hospital to heal. To monitor the healing process, they are asked to return to the hospital for rehabilitation  examination after 3, 6 and 12 months. Not all tests are applied for all patients in the study, since not all patients return on time for rehabilitation examination. Thus, the rehabilitation outcome scores also have a large amount of missing entries.

Based on the patient journey, we split these variables into two categories. The first one contains patient demographics and measurements realized during their stay at a hospital. These measurements include features such as age, BMI, blood tests of various chemicals related to tendon callus production, whether there was surgical intervention, or information on post-operative treatment. Variables in this category are referred to as \textit{predictors} in the following text. The second category is the \textit{scores}, and includes all metrics of rehabilitation outcomes such as ATRS or Foot and Ankle Outcome Score (FAOS). An example snapshot of the dataset is depicted in Figure \ref{dataset-snapshot}. In this work, we will impute the \textit{predictors} and predict the \textit{scores}.

\section{Methods}
\label{sec:methods}
We design an end-to-end probabilistic model to simultaneously impute the missing entries in the predictors and predict the rehabilitation outcomes. The data imputation part is a latent variable model which can be used separately or be part of the end-to-end model. For the prediction part, we provide multiple alternatives of modeling choices, including Bayesian linear regression and Bayesian neural networks, using either the learned latent variables or the imputed predictors as inputs. In this section, we first introduce the basic data imputation unit and then introduce our end-to-end model for simultaneous data imputation and rehabilitation outcome prediction.

\subsection{Measurement imputation}
\label{sec:imputation}
\begin{figure*}[t]
\centering   
\subfigure[Matrix factorization]{\label{fig:Methods-PMF}
\resizebox{0.15\textwidth}{!}{
\begin{tikzpicture}[scale=1]
\tikzstyle{halfobs} = [circle,draw=black,inner sep=1pt,
minimum size=20pt, font=\fontsize{10}{10}\selectfont, node distance=1]
  \node[halfobs]	(Pnm) {$P_{nm}$};
    \begin{scope}[on background layer]
      \fill[fill=gray!25] (Pnm.225) arc [start angle=225, end angle=405, radius=12pt];
    \end{scope}
  \node[latent, above=of Pnm, xshift=1.2cm] (u) {$u_{dn}$};
  \node[latent, above=of Pnm, xshift=-1.2cm]  (v) {$v_{dm}$};

  \edge {u,v} {Pnm} ; %

  \plate {D} {(v)(u)} {$D$} ;
  \plate {M} {(v)(Pnm) (D.north west)} {$M$} ;
  \plate {N} {(u)(Pnm) (M.north east) (D.north east) (M.south east)} {$N$} ;
  
\end{tikzpicture}}}%
\hspace{33pt}
\subfigure[Using dense predictors $\hat{\predictors}$]{\label{Methods:pgm:a}
\resizebox{0.26\textwidth}{!}{%
\begin{tikzpicture}[scale=1]
\tikzstyle{halfobs} = [circle,draw=black,inner sep=1pt,
minimum size=20pt, font=\fontsize{10}{10}\selectfont, node distance=1]
  \node[halfobs]                               (Pnm) {$P_{nm}$};
  \begin{scope}[on background layer]
      \fill[fill=gray!25] (Pnm.225) arc [start angle=225, end angle=405, radius=12pt];
    \end{scope}
  \node[latent, above=of Pnm, xshift=2.2cm] (u) {$u_{dn}$};
  \node[latent, above=of Pnm, xshift=-1.0cm]  (v) {$v_{dm}$};
  \node[halfobs, right=of Pnm, xshift=0.6cm]                               (Snp) {$S_{ns}$};
    \begin{scope}[on background layer]
      \fill[fill=gray!25] (Snp.225) arc [start angle=225, end angle=405, radius=10pt];
    \end{scope}
  \node[latent, right=of Snp, xshift=-0.6cm]  (b) {$b_s$};
  \node[latent, below=of Snp, xshift=-1.2cm]  (w) {$w_{ms}$};
 

  \edge {u,v} {Pnm} ; %
  \edge {Pnm} {Snp} ; %
    \edge {b} {Snp} ; %
  \edge {w} {Snp} ; %

  \plate {Anmv} {(v)(Pnm)(w)} {$M$} ;
  \plate {N} {(u)(Pnm)(Snp)(Anmv.north east)} {$N$} ;
  \plate {Bnpb} {(b)(Snp)(w)(Anmv.south east)} {$S$} ;
  \plate {D} {(v)(u)(Anmv.north west)(N.north west)} {$D$} ;

\end{tikzpicture}
}}%
\hspace{33pt}
\subfigure[Using patient traits $\hat{\mathbf{U}}$]{\label{Methods:pgm:b}
\resizebox{0.26\textwidth}{!}{%
\begin{tikzpicture}[scale=01]

\tikzstyle{halfobs} = [circle,draw=black,inner sep=1pt,
minimum size=20pt, font=\fontsize{10}{10}\selectfont, node distance=1]
  \node[halfobs]                               (Anm) {$P_{nm}$};
    \begin{scope}[on background layer]
      \fill[fill=gray!25] (Anm.225) arc [start angle=225, end angle=405, radius=12pt];
    \end{scope}
  \node[latent, above=of Anm, xshift=1.2cm] (u) {$u_{dn}$};
  \node[latent, above=of Anm, xshift=-1.2cm]  (v) {$v_{dm}$};
  \node[halfobs, right=of Anm, xshift=0.4cm] (Bnp) {$S_{ns}$};
    \begin{scope}[on background layer]
      \fill[fill=gray!25] (Bnp.225) arc [start angle=225, end angle=405, radius=11pt];
    \end{scope}
  \node[latent, right=of Bnp]  (b) {$b_s$};
  \node[latent, above=of b]  (w) {$w_{ds}$};
 
  \edge {u,v} {Anm} ; %
  \edge {b} {Bnp} ; %
  \edge {u} {Bnp} ; %
  \edge {w} {Bnp} ; %

  \plate {D} {(v)(u)(w)} {$D$} ;
  \plate {M} {(v)(Anm)(D.north west)} {$M$} ;
  \plate {P} {(b)(Bnp)(w)(D.north east)} {$S$} ;
  \plate {N} {(u)(Anm)(Bnp)(M.north east)(P.north west)(M.south east)(P.south west)} {$N$} ;

\end{tikzpicture}
}}
\caption{Graphical representation of the proposed models with probabilistic matrix factorization and Bayesian linear regression (or BNN). Half-shaded nodes describe partially observed variables. Panel 
(a) is the graphical representation of the probabilistic matrix factorization model for data imputation only. 
Panel (b) shows the model which uses the imputed measurements to predict the rehabilitation outcome.  
Panel (c) shows the model which uses the patient traits, a latent representation of the patient state, to predict the rehabilitation outcome.
}
\label{fig:Methods-EtE-LR-P}
\end{figure*}
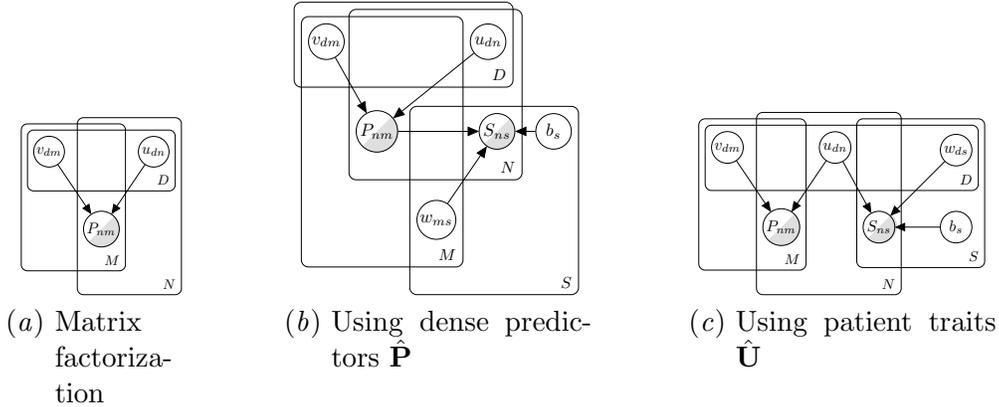

We first present the component of the model which aims to recover the missing measurements in the predictors part of the matrix, $\predictors \in  \mathbb{R}^{N \times M}$. We formulate the missing data imputation problem into a collaborative filtering problem. Typically, matrix factorization models are used in collaborative filtering for recommender systems. They work by decomposing the user-item interaction matrix into the product of two lower dimensionality rectangular matrices to predict unseen ratings. Thus, this technique can be used for data imputation.  We adopt a probabilistic matrix factorization based method \citep{matchbox}, where the latent traits are used to model the personal preference of users and the ratings for all items are predicted, but in this work we model the patient state and predict the missing measurements. The result is a latent variable model with Gaussian distributions, and its graphical representation is shown in Figure \ref{fig:Methods-PMF}.

For $N$ patients, $M$ measurements, $S$ scores and a latent space of size $D$, the model assumes that the patient measurement affinity matrix $\mathbf{A} \in \mathbb{R} ^{N \times M}$ is generated from the patient traits $\mathbf{U} \in \mathbb{R} ^{N \times D}$, which reflect the health status of the patient, and predictor traits $\mathbf{V} \in \mathbb{R}^{M \times D}$, which map different health status to measurements from various medical instruments. We use Gaussian distributions to model these entries. Thus, $p(\mathbf{U} | \sigma^2_\mathbf{U}) = \prod_{i = 1}^N \mathcal{N}(\mathbf{u}_i|\mu_\mathbf{U}, \sigma^2_\mathbf{U} \mathbf{1}), ~
p(\mathbf{V} | \sigma^2_\mathbf{V}) = \prod_{j = 1}^M \mathcal{N}(\mathbf{v}_j|\mu_\mathbf{V}, \sigma^2_\mathbf{V} \mathbf{1})$. Not all measurements are observed. The measurement imputation model is
\begin{equation}
	p(\predictors|\mathbf{U}, \mathbf{V}, \sigma^2_\predictors) = \prod_{n=1}^N \prod_{m=1}^M \Bigg[ \mathcal{N}(P_{nm} | \mathbf{u}_n^T \mathbf{v}_m, \sigma^2_\predictors) \Bigg]^{\mathbf{I}(n,m)},
\end{equation}
where $\mathcal{N}(x|\mu, \sigma^2)$ is the probability density function of the Gaussian distribution with mean $\mu$ and variance $\sigma^2$. $\mathbf{I}$ is an observation indication matrix, defined as 
\begin{eqnarray}
	\mathbf{I}(n,m) = \begin{cases}
		1 \text{~~if $P_{n,m}$ is observed,} & \\
		0 \text{~~otherwise.}
	\end{cases}
\end{eqnarray}
Thus, we can use the observed measurements to train the model, and use the generated measurements affinity $\mathbf{A}$ to impute the missing data. Eventually, we want the observed entries in $\mathbf{P}$ to be as close as possible to the corresponding entries in $\mathbf{A}$.

\subsection{Simultaneous data imputation and outcome prediction}
\label{sec:whole_model}
We present our proposed models which impute the missing entries and predict the rehabilitation outcome. Based on the model presented before, we add the second component to predict the scores matrix $\scores \in \mathbb{R}^{N \times S}$ using the patient information. The patient information can be the imputed measurement matrix as shown in Figure \ref{Methods:pgm:a}, or the patient trait vector which is a low-dimensional summary of the patient state as shown in Figure \ref{Methods:pgm:b}.

\paragraph{Bayesian linear regression.} We consider a Bayesian linear regression model first. The score is modeled as 
\begin{align}
	p(\scores \mid \mathbf{W}, \mathbf{b}, \mathbf{X}) &= \prod_{n=1}^N \prod_{s=1}^S \Bigg[ \mathcal{N}(S_{ns} \mid \mathbf{x}_n\mathbf{w}_s + b_s, \sigma_\scores^2) \Bigg]^{\mathbf{I'}_{ns}}\hspace{-0.1cm},\\ \nonumber
	p(\mathbf{W}) &= \mathcal{N}(\mathbf{W} \mid \mathbf{0}, \sigma_w^2 \mathbf{1})\,, \\ \nonumber
	p(\mathbf{b}) &= \mathcal{N}(\mathbf{b} \mid \mathbf{0}, \sigma_\mathbf{b}^2)\,,
\end{align}
where the input $\mathbf{X}$ is either the predictors or the patient traits, $\mathbf{W} \in \mathbb{R} ^{M \times S}$ and $\mathbf{b} \in \mathbb{R} ^S$ are weights and bias parameters for Bayesian linear regression. $\mathbf{S} \in \mathbb{R} ^{N \times S}$ indicates the observed rehabilitation scores, which can be seen as the rehabilitation outcome $\mathbf{B}$ masked by boolean observation indicator $\mathbf{I'}$, described similarly to the previous section. For a patient who has gone through the rehabilitation monitoring, our model can be used to predict the missing scores. For a new patient who has just received treatment, our model can predict the future healing outcome.
In the case of the predictors (Figure \ref{Methods:pgm:a}), we make use of the observed values so that 
the input $\mathbf{X}$ is 
$\hat{\mathbf{P}} = \mathbf{I} *\mathbf{P} + (1-\mathbf{I})*\mathbf{A}$,
where $\mathbf{I}$ is the $N \times M$ measurement observation indicator described in Section \ref{sec:imputation}.
In the case of the patient traits (Figure \ref{Methods:pgm:b}), we simply use $\hat{\mathbf{U}}$ as the input $\mathbf{X}$.
In fact, predictors $\hat{\predictors}$ contain more information but also more noise, and $\hat{\mathbf{U}}$ can be seen as a summary of each patient's characteristics. Therefore, we do our experiments with either $\hat{\predictors}$ or $\hat{\mathbf{U}}$ as inputs for the second component. Figure \ref{fig:Methods-EtE-LR-P} displays the graphical model in these two cases.

\paragraph{Bayesian neural network.}
We also consider a BNN, i.e. a neural network with probabilistic distributions on its weights and biases. In this case, we have the following conditional distribution of the scores
\begin{align}
p(\scores \mid \mathbf{\theta}, \mathbf{X}) = \prod_{n=1}^N \prod_{s=1}^S \Bigg[ \mathcal{N}(S_{ns} \mid \textbf{NN}(\mathbf{x}_n ; \mathbf{\theta}), \sigma_\scores^2) \Bigg]^{\mathbf{I'}_{ns}},
\end{align}
where $\mathbf{NN}$ is a Bayesian neural network parameterized by $\theta$, the collection of all weights and biases of the network. Typically, we consider fully connected layers with hyperbolic tangent activations. For a network of $L$ layers, we have
\begin{eqnarray}
	\mathbf{H}_{l} = \tanh\Bigl( \mathbf{H}_{l-1} \mathbf{W}_l + \mathbf{b}_l \Bigr) ~~~ \text{ for } l = 1,..., L\, ,
\end{eqnarray}
where $\mathbf{H}_l$ is the output of layer $l$ ($\mathbf{H}_0 = \mathbf{X}$), $\mathbf{W}_l$ is the matrix of weights from neurons of layer $l-1$ to neurons of layer $l$, and $\mathbf{b}_l$ is the bias vector for layer $l$. We start our experiments by setting up priors on weights according to Xavier's initialization \citep{glorot2010understanding}. That is, the prior variance of a weight $w$ that feeds into the $j$-th neuron of layer $l$ depends on $n_{lj, \text{in}}$, the number of neurons feeding into this neuron, and $n_{lj, \text{out}}$, the number of neurons which the result is fed to. For a weight $w_{lij}$ going from layer $l-1$, neuron $i$ to layer $l$, neuron $j$, we have
\begin{eqnarray}
p(w_{lij}) = \mathcal{N} (w_{lij} \mid 0, \sigma^2_{w_{lij}}), \\
\sigma^2_{w_{lij}} = \frac{2}{n_{lj, \text{in}} + n_{lj, \text{out}}}.
\end{eqnarray}
We limit the complexity of the networks that we evaluate a small number of hidden layers, since there is a limited amount \chedit{of} data. A model with increased complexity would be more prone to overfitting. The graphical model resembles the one in Figure \ref{fig:Methods-EtE-LR-P}, except that instead of the weights $\mathbf{W}$ and biases $\mathbf{b}$, we have the set of parameters of the network $\mathbf{\theta}$. The exact shape of the network is described in the experiments section.

\paragraph{Rehabilitation outcome prediction at various timestamps.} As we discussed before, the patient returns to hospital for rehabilitation monitoring after 3, 6, and 12 months. Thus at different timestamps, we have different amounts of observed data. We move further in the patient's journey and apply changes to the previous model so that it can be used at the 3-month or 6-month mark. In the first case, we rearrange our inputs and move the scores at 3 months from $\scores$ to $\predictors$. We denote these rearranged inputs $\predictors_3$ and $\scores_3$. We apply the same procedure for the 6-month mark and define $\predictors_6$ and $\scores_6$. We demonstrate in the next section that if we add more information from the healing monitoring, the performance of the final healing outcome prediction (at 12 months) is clearly improved.

\paragraph{Inference.} 
We run inference on this whole model in a end-to-end manner. We use variational inference with the KL divergence \citep{blei2017variational,zhang2017advances}.
We implemented all our models with the Edward library \citep{tran2016edward}, a probabilistic programming library. We use Gaussian distributions to approximate all posteriors.
\section{Experiments}
\label{sec:exp}

We evaluate our method in this section. We first verify our model and inference algorithm using a synthetic dataset. We then focus on the real-world ATR rehabilitation cohort and present preprocessing details. We compare our proposed method with multiple baselines. Finally, we discuss all experimental results. The experimental results show that our proposed end-to-end model clearly improves the predictive performance in comparison to the baselines. Additionally, we evaluate the rehabilitation outcome prediction at various timestamps and show that the accuracy of the rehabilitation outcome prediction increases with more observations.

\vspace{-5pt}
\subsection{Inference verification with synthetic data}
\label{sec:synthetic}

\begin{figure}
\floatbox[{\capbeside\thisfloatsetup{capbesideposition={right,top}}}]{figure}
{\caption{Synthetic data experiment. The plot demonstrates the learned latent traits with respect to the ground-truth. For a point in the figure, its x-coordinate is the true trait value, and y-coordinate is the corresponding recovered trait value. Different colors represent different latent traits. For brevity, we don't show all the recovered latent traits.}\label{Experiments:IV:U}}
{\begin{tikzpicture}
\tikzset{mark options={mark size=0.7}}
\begin{axis}[%
ylabel={Recovered traits},xlabel={True traits},
width=0.4\textwidth,height=0.4\textwidth,enlargelimits=false,
scatter/classes={%
    a={mark=*,draw=red,opacity=0.6},
    b={mark=*,draw=white!30!blue,opacity=0.6},
    c={mark=*,draw=green,opacity=0.6}}]
\addplot[scatter,only marks,%
    scatter src=explicit symbolic]%
table[meta=label] {
blue	red	label
1.14340843133	1.59854	a
1.85722448944	2.54657	a
-0.0999929671366	-0.102469	a
-0.151700293798	-0.174367	a
0.587284149696	0.831976	a
0.116887604291	0.143058	a
0.611347244799	0.89889	a
0.690348172293	1.01182	a
0.228770173518	0.358673	a
0.497018993971	0.719382	a
0.408581948016	0.568169	a
0.62073091161	1.00409	a
0.0115877861853	0.0157341	a
0.198925448217	0.320563	a
0.999270435857	1.4193	a
1.26178959821	1.74341	a
-0.193191512104	-0.139721	a
0.911613352088	1.24501	a
0.202619584686	0.288986	a
-0.160653492547	-0.143526	a
1.50151063973	2.06998	a
-0.168664529725	-0.157976	a
-0.236414161719	-0.265196	a
0.527143890818	0.751877	a
0.94867337023	1.39978	a
0.492567705325	0.750867	a
1.04478799052	1.50524	a
0.296552564784	0.529542	a
0.492447953986	0.790672	a
1.74372981128	2.34163	a
0.529292807795	0.743144	a
0.684121506113	0.947131	a
0.409569075063	0.688058	a
0.976024893454	1.40974	a
-0.0181959595618	0.124367	a
0.197396294628	0.384882	a
0.931602471539	1.26978	a
1.04990724496	1.5569	a
1.59336713906	2.21124	a
0.347974529881	0.636451	a
0.321423732069	0.495105	a
0.531971170039	0.853849	a
0.345510647097	0.448619	a
0.760879486833	0.968828	a
0.52789836376	0.836912	a
1.44454715904	2.00081	a
0.514006703869	0.731345	a
0.444322286056	0.670006	a
0.612355451192	0.863526	a
0.375276831298	0.523075	a
-0.179051860901	-0.187233	a
1.11197882443	1.45637	a
0.768627425257	1.07617	a
-0.946192289552	-1.24996	a
0.454953612928	0.614569	a
0.978077604828	1.41715	a
-0.293766654896	-0.271736	a
0.470329025744	0.702319	a
0.237635757157	0.285708	a
0.476055585889	0.676714	a
1.08777508258	1.43918	a
0.159646728267	0.240439	a
0.238814025725	0.377751	a
0.383556186731	0.557971	a
0.152674931908	0.260273	a
0.918149956259	1.28064	a
0.299665562259	0.434331	a
0.674852178701	0.982241	a
0.359012559393	0.489482	a
0.402697210935	0.678106	a
-0.543098016155	-0.645162	a
0.140986441534	0.299655	a
-0.460377769364	-0.562829	a
0.916464012885	1.29363	a
0.467262775953	0.66459	a
1.18694025143	1.54442	a
2.67182344582	3.49321	a
0.795380199699	1.05376	a
0.293035000615	0.4673	a
0.227186326108	0.401811	a
0.270454688247	0.379283	a
0.386621530754	0.760036	a
0.34611080944	0.50856	a
-0.123858934576	-0.15803	a
0.0476971086898	0.0949397	a
0.859294793909	1.25111	a
0.92187009255	1.20254	a
0.703248692709	0.969398	a
0.27407470417	0.535101	a
0.205266635332	0.293662	a
0.448096344541	0.651174	a
0.571029464837	0.858192	a
0.974883351261	1.3459	a
0.634693798177	0.789296	a
0.552020944821	0.768221	a
0.728996116794	1.01673	a
-0.716604059049	-0.865938	a
-0.690772460129	-0.844431	a
0.50403128415	0.741066	a
0.196669973991	0.384974	a
0.677665482491	1.07421	b
1.07738813888	1.63266	b
0.463859493852	0.662822	b
0.406943785813	0.650226	b
1.13572447281	1.70594	b
-0.536651742881	-0.740667	b
0.427833401406	0.659448	b
0.585018085739	0.793644	b
-0.177057994987	-0.219124	b
0.0798117047046	0.106607	b
-0.05749934748	-0.0714812	b
0.810968193854	1.19784	b
0.0724281988964	0.0663429	b
0.573089577409	0.865047	b
-0.0905494018146	-0.105435	b
0.747319319092	1.18108	b
0.892556129038	1.33827	b
0.330372593793	0.540612	b
-0.173788640594	-0.156613	b
0.427390401267	0.596765	b
0.677314912327	0.994945	b
0.253811552669	0.354863	b
0.577136144168	0.902029	b
0.273231977909	0.472869	b
0.792869515163	1.21291	b
0.959809402271	1.40074	b
0.724715205657	1.19875	b
0.926113301603	1.36233	b
1.13034608101	1.73255	b
0.209034469503	0.426782	b
0.757132740073	1.13585	b
0.55868984036	0.789523	b
1.67169188162	2.44663	b
1.14024845503	1.76807	b
0.628559314463	0.929656	b
2.01222131366	2.99936	b
0.506206574374	0.770954	b
0.514000674756	0.829275	b
0.524035896568	0.81511	b
1.5568536751	2.36527	b
0.0410923492672	0.102882	b
0.879842115493	1.36825	b
0.263964120132	0.472247	b
0.52855794066	0.851602	b
0.437418511312	0.572487	b
0.81089663105	1.24084	b
0.569403166704	0.835211	b
0.724976422112	1.10413	b
0.601960300165	0.807914	b
0.498263001432	0.678989	b
0.293513247171	0.509408	b
0.568638190848	0.866987	b
0.920697150813	1.36298	b
-0.558528678664	-0.772009	b
0.16377573466	0.216298	b
0.654814106516	1.06253	b
0.925786872611	1.4592	b
0.515598228665	0.755327	b
0.322014560169	0.498324	b
0.5030344693	0.76718	b
-0.367290860202	-0.533698	b
0.0615926790401	0.124238	b
0.624813993234	0.920134	b
0.339668798117	0.526491	b
0.571014861577	0.890287	b
0.333684678375	0.534078	b
0.134844318559	0.226767	b
0.344621171572	0.456936	b
0.0890922479564	0.177063	b
0.869244435215	1.31744	b
0.11543860975	0.20353	b
0.154274901685	0.245041	b
0.338444320868	0.524956	b
0.898482561766	1.32017	b
-0.0310065658743	0.0776547	b
-0.114817237686	-0.17039	b
-0.0834576299484	0.0499468	b
0.218614038518	0.44696	b
-0.112398440655	-0.153928	b
0.685026969007	1.05078	b
0.302300199258	0.426564	b
0.768292134505	1.19308	b
0.0516292304449	0.0796804	b
-0.310120156449	-0.507774	b
-0.0606290965196	-0.0995552	b
1.14463933427	1.73676	b
-0.0353480691156	0.103628	b
-0.0475521987994	-0.0600841	b
0.624119318691	0.926041	b
-0.0550619104165	-0.136963	b
0.470143885815	0.698906	b
0.636376604573	0.974019	b
-0.0486285763876	-0.0887386	b
-0.0760151521807	-0.0877707	b
0.467895496553	0.763885	b
0.12313970689	0.199553	b
0.296195369366	0.415265	b
0.778991354325	1.13282	b
0.41705680172	0.677681	b
0.637663959791	0.973862	b
0.646832358529	1.03395	c
-0.0642163628608	0.418967	c
0.815606102074	1.11321	c
0.708682977871	0.995606	c
0.421406088104	0.903137	c
0.608953136274	0.596058	c
0.565149336423	0.877842	c
-0.0280486355262	0.22348	c
0.0365817339917	0.0176195	c
0.409113404077	0.539689	c
0.524773997071	0.658093	c
1.12658049798	1.6038	c
0.837557352045	1.07693	c
0.410658173845	0.640249	c
0.320349608879	0.427479	c
0.644736630898	1.03229	c
0.674051940158	0.96161	c
0.935639417051	1.27318	c
0.61888100862	0.652714	c
0.241054580114	0.433499	c
-0.131377197613	0.162787	c
0.342631046382	0.451377	c
-0.285599882338	-0.257837	c
0.147215314085	0.292787	c
1.04266577223	1.53887	c
-0.0249792991016	0.262716	c
0.838819014309	1.22655	c
0.512260381894	0.866499	c
0.27376749265	0.64995	c
-0.226646173258	-0.048861	c
-0.0741763650364	0.217807	c
0.179498429839	0.447211	c
1.21313999331	1.95383	c
1.03790277319	1.70297	c
-0.0412683753656	0.0850291	c
0.0608385840833	0.608045	c
0.559938702111	0.961637	c
1.45117645031	1.9452	c
-0.52592061007	-0.307519	c
0.638312226457	1.19897	c
0.744623858981	0.886805	c
0.787766186864	1.16299	c
0.0987214239755	0.264619	c
0.108141005094	0.38796	c
0.612332642959	0.945167	c
-0.0158242807294	0.351324	c
-0.0546791931981	0.16439	c
0.14328099564	0.359938	c
0.155214668821	0.438244	c
0.212920877731	0.416051	c
0.913782950663	1.1018	c
-0.642913360819	-0.48651	c
0.647685598547	1.12124	c
-0.405127102202	-0.771903	c
1.63042549388	2.13603	c
0.337102501758	0.679732	c
0.0414056608871	0.193438	c
1.23561193303	1.62976	c
-0.990930347234	-1.06847	c
0.0348440773645	0.168944	c
0.495964858389	0.618563	c
0.442611455613	0.557304	c
-0.144692247574	0.0344598	c
0.380390674203	0.607043	c
0.544055102467	0.78695	c
0.972775066561	1.32402	c
0.997156432965	1.22354	c
-0.0675878647219	0.0917186	c
-0.0808210433311	-0.0520781	c
0.654527555784	1.00647	c
1.09037276722	1.25391	c
0.892068585238	1.0772	c
0.541453432439	0.657455	c
0.550949698696	0.968446	c
1.128641441	1.35199	c
0.281631131773	0.506985	c
0.958296419253	1.3635	c
-0.86002426204	-0.91729	c
1.36152043391	1.63726	c
-0.33431091033	-0.238768	c
0.316297785118	0.489012	c
1.56341227332	2.02047	c
0.00417907562722	0.0599443	c
-0.404807687344	-0.579239	c
0.386714320552	0.458281	c
0.723267372282	1.25145	c
0.87945403002	1.151	c
0.632329268422	0.91057	c
0.854619981019	1.1549	c
0.183199797017	0.209674	c
-0.245623240396	-0.147786	c
0.448642754862	0.788971	c
1.00368927149	1.33386	c
0.740027725653	0.962488	c
-0.0679102786236	0.111417	c
0.688964817906	0.903032	c
0.33264782182	0.357614	c
0.504983173041	0.762477	c
0.955897708108	1.27656	c
0.524181479087	0.742184	c
    };
\end{axis}
\end{tikzpicture}}
\end{figure}
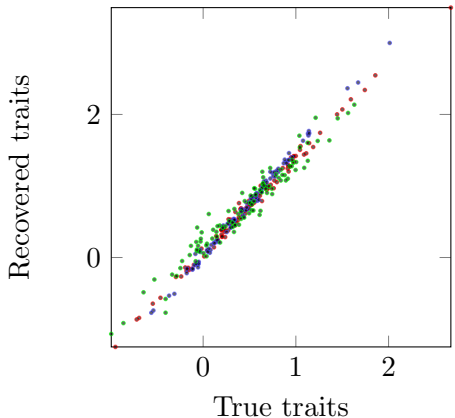

We test our model and inference algorithm with a synthetic dataset first. We build this synthetic dataset based on the generative process of the model and infer the latent parameters. We observe that our algorithms can successfully recover the latent parameters in all different settings. 
 
More precisely, we use $N = 100$, $M = 30$, $P = 10$, and a latent space of size $D = 10$. We generate the true patient and measurement traits by sampling from a normal distribution with mean 0.5 and variance 0.5. As discussed in \citep{groupbox}, this model is symmetric: parameters can rotate and inference can yield multiple valid solutions for $\mathbf{U}$ and $\mathbf{V}$. To ensure that we recover the true latent traits, we fix the upper square of $\mathbf{V}$ to the $D \times D$ identity matrix to avoid parameter rotation. Also, we add Gaussian noise with variance 0.1 to the resulting $\predictors$ matrix. We set priors matching this generative process. For evaluation, we randomly split the data into a training and a testing set with proportions 80\%-20\%. 

By comparing the inferred latent variables with their ground-truth values, we see that we can recover all of them. As an example, we show the ability of our end-to-end model with linear regression on $\predictors$ (Figure \ref{Methods:pgm:a}) to recover the patients' latent traits and to predict missing values in $\predictors$ and $\scores$. Figure \ref{Experiments:IV:U} depicts examples of the recovered patient traits. We can see that recovered values are close to true values because points in Figure \ref{Experiments:IV:U} are close to the diagonal of the square figure. Additionally, we evaluate the training and testing error with the Mean Absolute Error (MAE). We obtain an average training error of 0.042 for $\predictors$ and 0.027 for $\scores$, and an average testing error of 0.056 and 0.037 in the same order. These results validate our model and inference algorithm. Next, we evaluate our model on the real-life ATR dataset.

\subsection{Preprocessing of the Achilles tendon rupture rehabilitation cohort}
\label{sec:preprocessing} 
The cohort comes from clinical records with various formats, thus preprocessing is needed before we evaluate our method. First, we convert the whole dataset to numerical values, for example we convert the starting and ending time of the surgery to surgery duration. This process is done under the supervision of medical experts and the whole list of variables that we use are presented in the appendix. The ranges of measurements differ significantly due to the various units in use. We normalize every variable to be in the range of $[0,1]$. These are affine transformations and the original value can be easily recovered.
We do not fill in the missing data in the preprocessing steps as it is one of the goals of our model.

\subsection{Baselines}
\label{sec:baselines}
We compare our proposed method with seven variations of our proposed model with two types of baselines.
The first type of baseline uses traditional data imputation methods to impute the missing values in $\predictors$ and predict $\scores$. The second one is a two-stage version of our proposed model where data imputation and rehabilitation outcome prediction are performed in a sequential manner. 

\begin{table}[t]
\centering
\begin{tabular}{l c c c}
\toprule
  Component 2 Input & BLR  $\predictors$ &  BLR  $\scores$ &  BLR $\scores_{ATRS}$\\ 
\cmidrule(lr){1-4}
$\hat{\predictors}$ 2-stage (mean) 
& 0.228 $\pm$ 0.0014 & 0.230$\pm$ 0.008 & 0.200 $\pm$ 0.010
\\ 
$\hat{\predictors}$ 2-stage (OptSpace) 
& 0.224$\pm$ 0.0028 & 0.207 $\pm$ 0.009 & 0.193 $\pm$ 0.010 
\\ 
$\hat{\predictors}$ 2-stage (SoftImpute) 
& 0.2049 $\pm$ 0.002  &0.206 $\pm$ 0.008 &0.192 $\pm$ 0.010
\\ 
$\hat{\predictors}$ 2-stage (SVP) 
& 0.316 $\pm$ 0.003   & 0.205 $\pm$ 0.012   & 0.200 $\pm$ 0.014 
\\ 
$\hat{\predictors}$ 2-stage (IALM) 
& 0.237 $\pm$ 0.008 & 0.201 $\pm$ 0.011 & 0.201 $\pm$ 0.010 
\\ 
$\hat{\predictors}$ 2-stage (PMF) 
& 0.164 $\pm$ 0.002 & 0.220 $\pm$ 0.006 & 0.201 $\pm$ 0.007 
\\ 
$\hat{\mathbf{U}}$ 2-stage (PMF)
& 0.164 $\pm$ 0.002 & 0.237 $\pm$ 0.006 & 0.208 $\pm$ 0.006 
\\
$\hat{\predictors}~$ EE (proposed) 
& 0.181 $\pm$ 0.001   & 0.202$\pm$ 0.003 & 0.195$\pm$ 0.005 
\\
$\hat{\mathbf{U}}~$ EE (proposed) 
& 0.178$\pm$ 0.001 & \textbf{0.164}$\pm$ 0.004 & \textbf{0.146}$\pm$0.005 
\\
\bottomrule
\end{tabular}
\begin{tabular}{l c c c}
\toprule
  Component 2 Input & BNN  $\predictors$ &  BNN  $\scores$ &  BNN $\scores_{ATRS}$\\ 
\cmidrule(lr){1-4}
$\hat{\predictors}$ 2-stage (mean) 
& 0.228 $\pm$ 0.0014 & 0.233 $\pm$ 0.005 & 0.202 $\pm$ 0.005 \\ 
$\hat{\predictors}$ 2-stage (OptSpace) 
& 0.224$\pm$ 0.0028 & 0.203 $\pm$ 0.008 & 0.187$\pm$ 0.009  \\ 
$\hat{\predictors}$ 2-stage (SoftImpute) 
& 0.2049 $\pm$ 0.002   &0.201 $\pm$ 0.007 &0.186 $\pm$ 0.008 \\ 
$\hat{\predictors}$ 2-stage (SVP) 
& 0.316 $\pm$ 0.003   & 0.194 $\pm$ 0.010   & 0.187 $\pm$ 0.010 \\ 
$\hat{\predictors}$ 2-stage (IALM) 
& 0.237 $\pm$ 0.008 & 0.187$\pm$ 0.011 & 0.187 $\pm$ 0.009 \\ 
$\hat{\predictors}$ 2-stage (PMF) 
& 0.164 $\pm$ 0.002 & 0.207 $\pm$ 0.007 & 0.190 $\pm$ 0.007 \\ 
$\hat{\mathbf{U}}$ 2-stage (PMF)
& 0.164 $\pm$ 0.002 & 0.208$\pm$ 0.007 & 0.190 $\pm$ 0.007\\
$\hat{\predictors}~$ EE (proposed) 
& 0.158 $\pm$ 0.001   & \textbf{0.152}$\pm$0.004& \textbf{0.143}$\pm$0.004\\
$\hat{\mathbf{U}}~$ EE (proposed) 
& 0.167$\pm$0.001 & 0.174$\pm$0.003 & 0.152$\pm$ 0.003 \\
\bottomrule
\end{tabular}
\caption{ Mean Absolute Error (MAE) and standard deviation over 5 runs for outcome prediction. Each time, we use random splits of the data with 80\% data for training and 20\%  data for testing. ``EE'' indicates end-to-end which is our proposed model. ``2-stage'' is the baseline model where data imputation and rehabilitation outcome prediction are performed in a sequential manner. BLR stands for Bayesian Linear Regression and BNN stands for Bayesian Neural Network. 
For the 2-stage models, the error on $\predictors$ remains the same because the matrix $\predictors$ is imputed once with the mean imputation or the matrix factorization based methods. In addition, we report the MAE for the ATRS separately. The target is that the MAE of $\scores_{ATRS}$ gets smaller than $0.1$, because only a difference larger than $0.1$ is considered to be clinical different.
\vspace{-5pt}
}
\label{Results}
\end{table}

\paragraph{Traditional data imputation.}
We first consider imputing the per-patient mean \citep{scheffer2002dealing} to all missing values. For each patient, the mean value of their observations belonging to the training set is imputed to all their missing measurements. We also apply traditional matrix factorization based methods: OptSpace \citep{keshavan2010matrix}, SoftImpute \citep{mazumder2010spectral},
 Singular Value Projection \citep{jain2010guaranteed} and Inexact Augmented Lagrange Multiplier \citep{lin2010augmented}, to missing values. The predicted values based on observations are imputed to all the missing values. We then use the imputed data to predict rehabilitation outcomes using Bayesian linear regression and Bayesian neural network.

\paragraph{Two-stage version of the proposed model.}
We run inference on the probabilistic matrix factorization part and only retrieve predictions for the first part of the dataset, $\hat{\predictors}$, and the patient trait matrix $\hat{\mathbf{U}}$. Then, we define the second model which uses either linear regression or a neural network on this output to give the scores predictions, $\hat{\scores}$. Inference is run separately on each component. The intent is to compare our end-to-end model with its direct multi-staged equivalent. 

\subsection{Results}
\label{sec:results}
We split the training and testing set to reflect the treatment journey. 
In all of our experiments, we first pick training and testing data for $\predictors$ and $\scores$ with the following strategy: we randomly pick 80\% of the patients, take all their available data for training and leave the remaining 20\% of patients for testing. In other words, we split the dataset on a per-patient basis. We do this since the goal of our work is to predict the rehabilitation outcome after a patient receives the initial treatment. All experiments are repeated 5 times, with all the learned variables getting reset at each run.

We use grid search for hyper-parameter tuning, starting with the matrix factorization part. We observe that the prior mean on the traits has little effect on the end performance. However, the prior variance on both traits and scores has a big impact on how the model fits the training data. We evaluate latent space sizes $D \in [1, 20]$ as well as latent trait variances $\sigma^2_{\mathbf{U}}$ and $\sigma^2_{\mathbf{V}}$, ranging from 0.1 to 0.9 by steps of 0.2. We find the optimal $D$ to be 8 and the optimal variance to be 0.5.  Next, we tune the linear regression and neural network. We start by tuning the linear regression then turn it into a neural network with growing complexity by adding activation functions and layers as soon as we find that the model lacks expressive power. We run grid searches on the prior means and variances of the weights and biases as well as the observation noise on $\scores$. We notice that for the model to properly fit the training data, weights need to have a very small variance. This is expected since the data is very high-dimensional and the results of dot products need to be constrained in $[0, 1]$). Doing so, we find that dividing the weight variance computed with Xavier's initialization by $10^3$ and the prior observation noise by $10^4$ yields the best results. 
 
We report the performance of the rehabilitation outcome prediction in Table \ref{Results}. The Mean Absolute Error (MAE) on the testing set is used as metric for our results. In practice, only a difference larger than $0.1$ is considered to be clinically different. Thus, results with MAE within $0.1$ are ideal.
To compare with this standard, we evaluate prediction methods only with 11 ATRS (10 criteria and the sum), whose results are shown in $\scores_{ATRS}$ columns of Table \ref{Results} and Table \ref{Results-EPM}. We can see that our proposed end-to-end model with neural network applied on the whole $\predictors$ matrix achieves the best performance for predicting rehabilitation outcomes and its $\scores_{ATRS}$ result is close to the ideal MAE target 0.1. We see that for predicting $\scores$, using the patient traits $\hat{\mathbf{U}}$ works better in the case of linear regression, and using the whole predictors matrix $\hat{\predictors}$ works better in the case of neural network. This is certainly due to the fact that the dimensionality of $\hat{\predictors}$ makes it difficult for a simple model such as linear regression to extract the key features; a task that a more complex neural network would manage better. In these experiments, we start with a neural network that basically replicates the linear regression then gradually add complexity until we can't improve the performance without overfitting. The optimal network we found has 1 hidden layer with $P$ (the number of columns of $\scores$) hidden units and a hyperbolic tangent activation.

Our proposed method shows clear improvement on the rehabilitation outcome prediction over baselines. We can also see that latent variable models have a good performance on the missing value imputation. Our proposed model is trained for the rehabilitation outcome prediction, so SVP and IALM could have the better performance on the missing value imputation than ours.

\begin{table}[th]
\centering
\begin{tabular}{l c c c}
\toprule
& Discharge $\hat{\mathbf{P}}$ & 3 Month $ \hat{\mathbf{P_3}}$ & 6 Month $\hat{\mathbf{P_6}}$ \\
  \cmidrule(lr){1-4}
  MAE $\scores_{3}$  		&  0.177 $\pm$ 	0.006&& \\
  MAE $\scores_{ATRS-3}$  	& 0.173 $\pm$ 0.005&& \\
  MAE $\scores_{6}$ 		&0.172$\pm$ 0.007	& 0.178$\pm$ 0.006&\\
  MAE $\scores_{ATRS-6}$ & 0.167 $\pm$ 0.009 & 0.169$\pm$0.010 	&\\
MAE $\scores_{12}$ & 0.138 $\pm$ 0.006 & 0.140 $\pm$ 0.006 & \textbf{0.132}$\pm$ 0.006 \\
   MAE $\scores_{ATRS-12}$ & 0.111 $\pm$ 0.003 	& 0.114 $\pm$ 0.004& \textbf{0.108}$\pm 0.003$\\ 
\bottomrule
\end{tabular}
\caption{Rehabilitation outcome prediction performance comparison at various timestamps. Our proposed model with Bayesian neural network is used for this evaluation. We show that the final rehabilitation outcome prediction accuracy increases with time and our model can be used for the rehabilitation outcome prediction at various rehabilitation stages. $\scores_{ATRS}$ is evaluated only with ATRS.
\vspace{-5pt}
}
\label{Results-EPM}
\end{table}

\paragraph{Evaluation of the rehabilitation outcome prediction at different timestamps.} 
Here we evaluate the ability of our model to predict scores at different timestamps when we extend $\predictors$ to include scores at 3 months (yielding $\predictors_3$) and 6 months (yielding $\predictors_6$). We report the performance per-timestamp of our model with $\predictors$, $\predictors_3$ and $\predictors_6$ in Table \ref{Results-EPM}. 

We observe that including future measurements helps predicting the final scores. Including all the previously observed data in the predictors helps improving the accuracy of future score predictions. Moreover, results of the ATRS prediction at 12 months are close to 0.1 which is our target value.

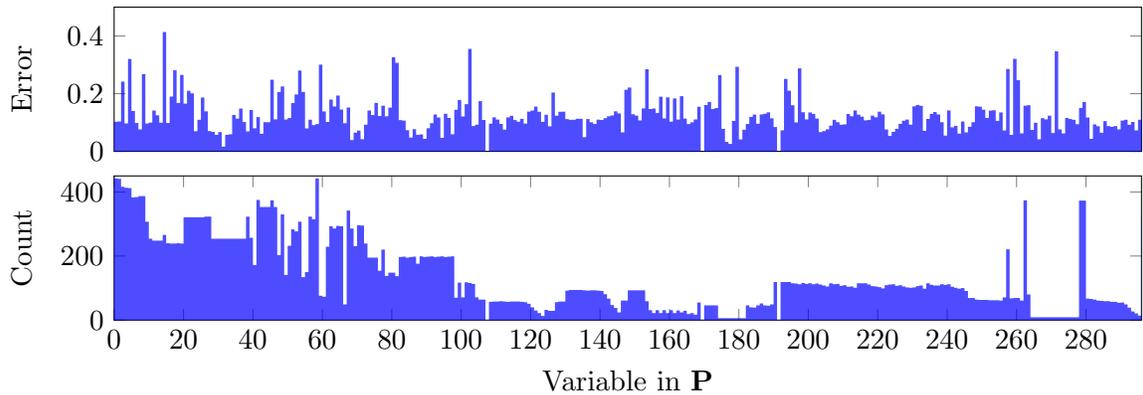
\begin{figure}[th]
	\centering
	\begin{tikzpicture}
\begin{axis}[ylabel={Error},width=\textwidth,height=3.5cm,ymin=0,ymax=0.5,enlargelimits=false,xtick style={draw=none},xtick=\empty]
\addplot[const plot,fill=blue,draw=none,fill opacity=0.7] 
coordinates
{
(0, 0.102686709408)
(1, 0.103682123389)
(2, 0.241859940079)
(3, 0.0968578847272)
(4, 0.320377860723)
(5, 0.140016380595)
(6, 0.0984266881304)
(7, 0.0761137004131)
(8, 0.268125572273)
(9, 0.096585369174)
(10, 0.0997915729541)
(11, 0.14164351758)
(12, 0.124989810124)
(13, 0.0993840035055)
(14, 0.41388746496)
(15, 0.098404332432)
(16, 0.189998473084)
(17, 0.281050402111)
(18, 0.1679569679)
(19, 0.265079045695)
(20, 0.165546449262)
(21, 0.209983389053)
(22, 0.201251678374)
(23, 0.0698366250476)
(24, 0.10948859526)
(25, 0.18668590804)
(26, 0.138702655417)
(27, 0.0703623010117)
(28, 0.0674528520952)
(29, 0.0575080610149)
(30, 0.0670889349613)
(31, 0.0162941329521)
(32, 0.0573978319089)
(33, 0.0592651390632)
(34, 0.126242154034)
(35, 0.113565838535)
(36, 0.148394254854)
(37, 0.101850164215)
(38, 0.0693135806738)
(39, 0.143475598015)
(40, 0.0799806427818)
(41, 0.118720611829)
(42, 0.0610842699436)
(43, 0.101288873199)
(44, 0.101158925185)
(45, 0.248301877345)
(46, 0.110979791584)
(47, 0.205524656699)
(48, 0.225094427135)
(49, 0.110463521039)
(50, 0.115684507385)
(51, 0.16761626561)
(52, 0.1972694286)
(53, 0.280305641417)
(54, 0.206088030911)
(55, 0.079623538057)
(56, 0.109721523574)
(57, 0.0909502209628)
(58, 0.094701770371)
(59, 0.300680672848)
(60, 0.138493832467)
(61, 0.102115158984)
(62, 0.179617622859)
(63, 0.155616009122)
(64, 0.193550814248)
(65, 0.14497558991)
(66, 0.0975615901407)
(67, 0.1520055774)
(68, 0.0392506329946)
(69, 0.0648144900867)
(70, 0.0720896808422)
(71, 0.0423214628833)
(72, 0.0923713504145)
(73, 0.141773834163)
(74, 0.12591412193)
(75, 0.168038131522)
(76, 0.122499124764)
(77, 0.158431469229)
(78, 0.121663167929)
(79, 0.151636510685)
(80, 0.32639678944)
(81, 0.306996607305)
(82, 0.108354881545)
(83, 0.106614254043)
(84, 0.073534556106)
(85, 0.0477911534257)
(86, 0.076894043012)
(87, 0.0571604540376)
(88, 0.0587881696179)
(89, 0.0433918615457)
(90, 0.0803436662464)
(91, 0.0979330372304)
(92, 0.127859504684)
(93, 0.117437546235)
(94, 0.0469516958653)
(95, 0.130413970476)
(96, 0.116006365316)
(97, 0.0592740646632)
(98, 0.144225255021)
(99, 0.178180992001)
(100, 0.120989304983)
(101, 0.163671168595)
(102, 0.35527231236)
(103, 0.0877787272851)
(104, 0.0926707457404)
(105, 0.174297196299)
(106, 0.108889219953)
(107, 0)
(108, 0.0962465414861)
(109, 0.118032595687)
(110, 0.110287633386)
(111, 0.0951494424273)
(112, 0.0753309460669)
(113, 0.117718580159)
(114, 0.123263817338)
(115, 0.108628366191)
(116, 0.102956627528)
(117, 0.112655514607)
(118, 0.0963077383051)
(119, 0.13788482717)
(120, 0.14159626474)
(121, 0.155145790483)
(122, 0.137190391904)
(123, 0.102410874189)
(124, 0.127078709242)
(125, 0.0868122692333)
(126, 0.204010182237)
(127, 0.123714038895)
(128, 0.137199391559)
(129, 0.137723823891)
(130, 0.111926042796)
(131, 0.10942100954)
(132, 0.109706986794)
(133, 0.113163220802)
(134, 0.112864991092)
(135, 0.0492826099701)
(136, 0.111998415313)
(137, 0.100836595673)
(138, 0.0949580984655)
(139, 0.110027496155)
(140, 0.10876234853)
(141, 0.115005995654)
(142, 0.121745692197)
(143, 0.124528845674)
(144, 0.138807915268)
(145, 0.131419189757)
(146, 0.0659071220088)
(147, 0.213528012338)
(148, 0.221299912874)
(149, 0.128721878796)
(150, 0.123923042339)
(151, 0.107414446025)
(152, 0.148894466263)
(153, 0.284728326819)
(154, 0.14679042024)
(155, 0.148452609744)
(156, 0.133309423058)
(157, 0.189496024873)
(158, 0.114282373219)
(159, 0.187091069571)
(160, 0.103029455732)
(161, 0.183491767933)
(162, 0.110359006579)
(163, 0.191451262133)
(164, 0.11368158917)
(165, 0.0946740366551)
(166, 0.100709913965)
(167, 0.110695197192)
(168, 0.154754708843)
(169, 0)
(170, 0.160663917378)
(171, 0.171459732615)
(172, 0.146747445758)
(173, 0.150127476648)
(174, 0.264768893763)
(175, 0.0781284027749)
(176, 0.0328564894372)
(177, 0.0258343145012)
(178, 0.105474883759)
(179, 0.293278475551)
(180, 0.0414881591211)
(181, 0.0750235135808)
(182, 0.0949888817087)
(183, 0.119403571635)
(184, 0.127608224028)
(185, 0.0958701998148)
(186, 0.127950681102)
(187, 0.130362032879)
(188, 0.13545889285)
(189, 0.114545569151)
(190, 0.0844230952866)
(191, 0)
(192, 0.0729428194595)
(193, 0.251332660138)
(194, 0.210334509052)
(195, 0.159942172218)
(196, 0.100530425917)
(197, 0.287930303473)
(198, 0.135762297612)
(199, 0.110794442471)
(200, 0.134878248352)
(201, 0.130109327264)
(202, 0.113767730579)
(203, 0.0667952681637)
(204, 0.0703723698779)
(205, 0.0763503936692)
(206, 0.0893252521948)
(207, 0.109243068805)
(208, 0.101173964756)
(209, 0.0923027071527)
(210, 0.0937423818218)
(211, 0.129715326332)
(212, 0.121808264728)
(213, 0.141692333501)
(214, 0.127367136871)
(215, 0.120184525914)
(216, 0.107662570473)
(217, 0.108789070994)
(218, 0.124145965458)
(219, 0.118247022667)
(220, 0.137633872835)
(221, 0.128345620932)
(222, 0.0760227709885)
(223, 0.0535183998643)
(224, 0.0677201767118)
(225, 0.0782886536724)
(226, 0.0954225325901)
(227, 0.106283506833)
(228, 0.110255464077)
(229, 0.0926809054759)
(230, 0.156177731635)
(231, 0.160911961346)
(232, 0.158000844805)
(233, 0.0713309736726)
(234, 0.112607624849)
(235, 0.130126672385)
(236, 0.136402599271)
(237, 0.128056204169)
(238, 0.101205394483)
(239, 0.0546955731804)
(240, 0.141879052607)
(241, 0.0834029822877)
(242, 0.0885524640354)
(243, 0.0614579099486)
(244, 0.103508643737)
(245, 0.0659216890618)
(246, 0.0852894396429)
(247, 0.101156609144)
(248, 0.156251920875)
(249, 0.154708740967)
(250, 0.144570876757)
(251, 0.115898750348)
(252, 0.141407739688)
(253, 0.142507660883)
(254, 0.10584346925)
(255, 0.134920533979)
(256, 0.0725637033687)
(257, 0.28564101867)
(258, 0.057135595995)
(259, 0.320913658261)
(260, 0.246791224373)
(261, 0.0623889497132)
(262, 0.158423173883)
(263, 0.160077750192)
(264, 0.0743328351699)
(265, 0.0994047495065)
(266, 0.0419098863153)
(267, 0.1151987019)
(268, 0.110957365052)
(269, 0.123832606238)
(270, 0.0636130085455)
(271, 0.34683235787)
(272, 0.0761598613304)
(273, 0.0621993270191)
(274, 0.116047500907)
(275, 0.112197869863)
(276, 0.109345297656)
(277, 0.0942129878006)
(278, 0.150713478649)
(279, 0.171435138104)
(280, 0.117348185179)
(281, 0.0434753732943)
(282, 0.0936145417862)
(283, 0.0871538869463)
(284, 0.0646083908209)
(285, 0.104738536437)
(286, 0.0875759080101)
(287, 0.0852238867368)
(288, 0.0876786943709)
(289, 0.0763330992169)
(290, 0.106765242143)
(291, 0.10959634314)
(292, 0.0921465567899)
(293, 0.102053072645)
(294, 0.0727388800868)
(295, 0.109619186486)
(296, 0.183913550974)
} \closedcycle;
\end{axis}

\end{tikzpicture}

\begin{tikzpicture}

\begin{axis}[ylabel={Count}, xlabel={Variable in $\predictors$},width=\textwidth,height=3.5cm,ymin=0,ymax=450,enlargelimits=false,]
\addplot[const plot,fill=blue,draw=none, opacity=0.7] 
coordinates
{
(0, 442)
(1, 441)
(2, 416)
(3, 413)
(4, 412)
(5, 383)
(6, 384)
(7, 387)
(8, 387)
(9, 307)
(10, 254)
(11, 248)
(12, 248)
(13, 248)
(14, 266)
(15, 240)
(16, 239)
(17, 239)
(18, 240)
(19, 239)
(20, 321)
(21, 321)
(22, 321)
(23, 321)
(24, 321)
(25, 321)
(26, 323)
(27, 323)
(28, 254)
(29, 254)
(30, 254)
(31, 254)
(32, 254)
(33, 254)
(34, 254)
(35, 254)
(36, 254)
(37, 254)
(38, 323)
(39, 257)
(40, 172)
(41, 375)
(42, 353)
(43, 353)
(44, 353)
(45, 374)
(46, 353)
(47, 203)
(48, 330)
(49, 141)
(50, 232)
(51, 283)
(52, 277)
(53, 307)
(54, 134)
(55, 150)
(56, 323)
(57, 315)
(58, 442)
(59, 76)
(60, 73)
(61, 229)
(62, 293)
(63, 286)
(64, 294)
(65, 293)
(66, 49)
(67, 342)
(68, 286)
(69, 231)
(70, 296)
(71, 295)
(72, 239)
(73, 195)
(74, 195)
(75, 195)
(76, 154)
(77, 220)
(78, 138)
(79, 148)
(80, 148)
(81, 137)
(82, 197)
(83, 198)
(84, 195)
(85, 197)
(86, 198)
(87, 176)
(88, 199)
(89, 197)
(90, 198)
(91, 199)
(92, 197)
(93, 198)
(94, 199)
(95, 197)
(96, 198)
(97, 199)
(98, 70)
(99, 117)
(100, 71)
(101, 118)
(102, 116)
(103, 113)
(104, 71)
(105, 64)
(106, 64)
(107, 0)
(108, 57)
(109, 58)
(110, 58)
(111, 59)
(112, 58)
(113, 57)
(114, 58)
(115, 58)
(116, 58)
(117, 57)
(118, 56)
(119, 51)
(120, 40)
(121, 30)
(122, 22)
(123, 13)
(124, 32)
(125, 29)
(126, 29)
(127, 56)
(128, 58)
(129, 58)
(130, 92)
(131, 94)
(132, 94)
(133, 94)
(134, 94)
(135, 93)
(136, 92)
(137, 93)
(138, 93)
(139, 92)
(140, 91)
(141, 81)
(142, 67)
(143, 47)
(144, 38)
(145, 24)
(146, 61)
(147, 61)
(148, 93)
(149, 93)
(150, 93)
(151, 93)
(152, 93)
(153, 59)
(154, 31)
(155, 22)
(156, 31)
(157, 22)
(158, 31)
(159, 22)
(160, 31)
(161, 22)
(162, 30)
(163, 24)
(164, 29)
(165, 19)
(166, 22)
(167, 17)
(168, 55)
(169, 0)
(170, 46)
(171, 46)
(172, 46)
(173, 46)
(174, 6)
(175, 6)
(176, 6)
(177, 6)
(178, 6)
(179, 6)
(180, 6)
(181, 6)
(182, 46)
(183, 40)
(184, 39)
(185, 46)
(186, 52)
(187, 46)
(188, 45)
(189, 49)
(190, 119)
(191, 0)
(192, 119)
(193, 119)
(194, 119)
(195, 115)
(196, 114)
(197, 112)
(198, 116)
(199, 113)
(200, 115)
(201, 112)
(202, 114)
(203, 111)
(204, 108)
(205, 105)
(206, 115)
(207, 112)
(208, 113)
(209, 107)
(210, 109)
(211, 104)
(212, 105)
(213, 101)
(214, 115)
(215, 115)
(216, 115)
(217, 111)
(218, 106)
(219, 105)
(220, 103)
(221, 99)
(222, 109)
(223, 111)
(224, 112)
(225, 107)
(226, 109)
(227, 105)
(228, 102)
(229, 101)
(230, 102)
(231, 106)
(232, 108)
(233, 98)
(234, 115)
(235, 110)
(236, 108)
(237, 108)
(238, 112)
(239, 110)
(240, 112)
(241, 104)
(242, 101)
(243, 98)
(244, 100)
(245, 91)
(246, 69)
(247, 69)
(248, 64)
(249, 63)
(250, 63)
(251, 63)
(252, 62)
(253, 62)
(254, 62)
(255, 61)
(256, 71)
(257, 221)
(258, 68)
(259, 69)
(260, 69)
(261, 61)
(262, 374)
(263, 80)
(264, 9)
(265, 9)
(266, 9)
(267, 9)
(268, 9)
(269, 9)
(270, 9)
(271, 9)
(272, 9)
(273, 9)
(274, 9)
(275, 9)
(276, 9)
(277, 9)
(278, 373)
(279, 373)
(280, 67)
(281, 66)
(282, 64)
(283, 61)
(284, 60)
(285, 59)
(286, 59)
(287, 57)
(288, 57)
(289, 54)
(290, 55)
(291, 49)
(292, 39)
(293, 27)
(294, 21)
(295, 14)
(296, 323)
} \closedcycle;
\end{axis}
\end{tikzpicture}
	\vspace{-5pt}
    \caption{
    Per-variable mean MAE and number of data points available for training for the predictors $\predictors$.
    \vspace{-5pt}
    }
    \label{Experiments-PVP}
\end{figure}

\begin{figure}[th]
	\centering
	\begin{tikzpicture}
\begin{axis}[ylabel={Error},width=\textwidth,height=3.5cm,ymin=0,ymax=0.5,enlargelimits=false,
ytick={0.0,0.1,0.2,0.3,0.4,0.5},
yticklabel style={/pgf/number format/.cd,fixed,precision=1}
,xtick style={draw=none}]
\addplot[const plot,fill=blue,draw=none,fill opacity=0.7] 
coordinates
{
(0,0.1940256420771281)(1,0.2136600520213445)(2,0.22136102279027306)(3,0.2039252523581187)(4,0.2052365026871363)(5,0.21963735604286191)(6,0.2126018657684326)(7,0.13486311752349142)(8,0.14226469801117977)(9,0.22069868793090186)(10,0.15363345539616974)(11,0)
(16,0.2594263583421707)(17,0.31066280901432036)(18,0.22328151226043702)(19,0.24878281984064313)(20,0.23355063080787658)(21,0.28963599953436314)(22,0.2073553933037652)(23,0.29710982536014763)(24,0.28183955318397946)(25,0.19061995995672126)(26,0.22168594718717566)(27,0)
(35,0.19702406370320083)(36,0.1701438236282246)(37,0.17446012003668426)(38,0.13543331972483932)(39,0.17237849996007723)(40,0.1615825554780793)(41,0.14891460195154704)(42,0.19416277324944212)(43,0.20682191629158825)(44,0.14752030686328285)(45,0.15202419269821027)(46,0.3103282716633244)(47,0.4130007880821563)(48,0)
(53,0.20032184393191013)(54,0.22189672348987785)(55,0.21670085683433107)(56,0.18387758979311972)(57,0)
(58,0.2898155185911391)(59,0)
} \closedcycle;
\addplot[const plot,fill=red,draw=none,fill opacity=0.6] 
coordinates
{
(11,0.26687063643897796)(12,0.26805447367416024)(13,0.38996327207202003)(14,0.21205383645163645)(15,0.20636978474530307)(16,0)
(27,0.24302766369838344)(28,0.31977772933465465)(29,0.27657218645977716)(30,0.35820808234038176)(31,0.37114880347448115)(32,0)
(48,0.15501973653369475)(49,0.13076181098595582)(50,0.11043064012493287)(51,0.13798739489387063)(52,0.1770238626533085)(53,0)
(57,0.29352138292508173)(58,0)

} \closedcycle;
\addplot[const plot,fill=black!30!green,draw=none,fill opacity=0.7] 
coordinates
{
(32,0.2393633914382561)(33,0.19497177873442811)(34,0.1772906164328257)(35,0)
(59,0.47668843981089537)(60,0.15918259069501425)(61,0.22338567946546822)(62,0.30470215255378663)(63,0)

} \closedcycle;
\draw[thick,color=red] (axis cs:0,0.1) -- (axis cs:63,0.1);
\end{axis}
\end{tikzpicture}


\begin{tikzpicture}
\begin{axis}[ylabel={Count}, 
width=\textwidth,height=3.5cm,ymin=0,ymax=100,enlargelimits=false,]
\addplot[const plot,fill=blue,draw=none, opacity=0.7] 
coordinates
{
(0,30)(1,30)(2,30)(3,30)(4,30)(5,25)(6,25)(7,24)(8,24)(9,24)(10,23)(11,0)
(16,20)(17,20)(18,20)(19,20)(20,20)(21,19)(22,19)(23,19)(24,18)(25,19)(26,19)(27,0)
(35,58)(36,58)(37,58)(38,58)(39,58)(40,57)(41,57)(42,57)(43,57)(44,57)(45,58)(46,88)(47,57)(48,0)
(53,57)(54,57)(55,56)(56,56)(57,0)
(58,54)(59,0)
} \closedcycle;
\addplot[const plot,fill=red,draw=none, opacity=0.6] 
coordinates
{
(11,3)(12,3)(13,3)(14,3)(15,3)(16,0)
(27,3)(28,3)(29,3)(30,3)(31,3)(32,0)
(48,45)(49,45)(50,45)(51,45)(52,45)(53,0)
(57,43)(58,0)
} \closedcycle;
\addplot[const plot,fill=black!30!green,draw=none, opacity=0.7] 
coordinates
{
(32,4)(33,4)(34,3)(35,0)
(59,14)(60,9)(61,9)(62,7)(63,0)
} \closedcycle;
\end{axis}

\draw [
decorate,
decoration={brace,amplitude=10pt,mirror},
xshift=0pt,
yshift=0pt] 
(0,-0.5) -- (15*0.23,-0.5) 
node [black,midway,xshift=0cm,yshift=-0.5cm] {3 months};

\draw [
decorate,
decoration={brace,amplitude=10pt,mirror},
xshift=0pt,
yshift=0pt] 
(15*0.23,-0.5) -- (33*0.23,-0.5) 
node [black,midway,xshift=0cm,yshift=-0.5cm] {6 months};
\draw [
decorate,
decoration={brace,amplitude=10pt,mirror},
xshift=0pt,
yshift=0pt] 
(33*0.23,-0.5) -- (60*0.23,-0.5) 
node [black,midway,xshift=0cm,yshift=-0.5cm] {12 months};
\end{tikzpicture}
	\vspace{-5pt}
    \caption{Per-variable mean absolute error (MAE) and number of data points available for training for different components of scores at different rehabilitation time. Blue bars represent ATRS. Red bars represent FAOS. Green bars represent other test scores.
    \vspace{-10pt}
    }
    \label{Experiments-PVS}
\end{figure}
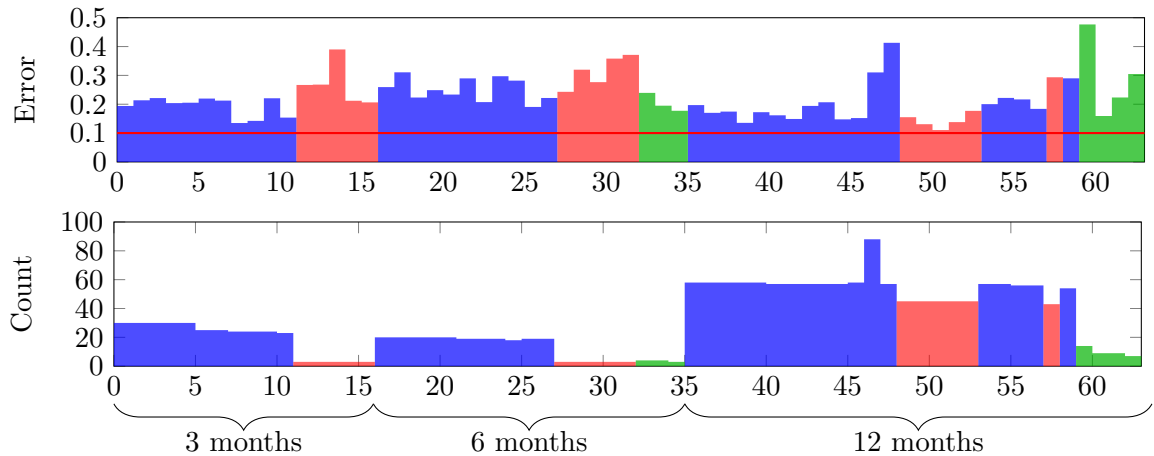

\paragraph{Per-variable analysis.}
We further evaluate the prediction accuracy of our best performing model by looking the mean error for each variable. Taking the model with $\hat{\mathbf{P}}$ with BNN as an example, Figures \ref{Experiments-PVP} and \ref{Experiments-PVS} display the errors and the number of data points available for each variable. 

In Figure \ref{Experiments-PVS}, we show that the number of scores per period varies. In fact, each period has at least 11 ATRS (10 criteria and the sum in blue) and 5 FAOS scores (in red). On top of that, scores at 6 and 12 months both include additional tests such as the evaluation of the heel rise angle (in green). The clinical practice uses scores at 12 months more because they can reflect rehabilitation states better. Figure \ref{Experiments-PVS} shows that our model is able to predict the rehabilitation outcome at 12 months better comparing to 3 and 6 months.

\section{Conclusions} 
We developed a probabilistic end-to-end framework to simultaneously predict the rehabilitation outcome and impute the missing entries in data cohort in the context of Achilles Tendon Rupture (ATR) rehabilitation. We evaluated our model and compared its performance with multiple baselines. We demonstrated a clear improvement in the accuracy of the predicted outcomes in comparison with traditional data imputation methods. Additionally, the performance of our method on rehabilitation outcome prediction is close to the ideal clinical result.

\paragraph{Future work.}
There is still considerable work to be done in the interpretation of our results, in a clinical sense. An analysis of the impact of each predictor in each model as in \citet{popkes2019interpretable} and a discussion on how these relate to the ATR clinical experiences are desirable to strengthen this work from a medical point of view.
Additionally, we are keen to work closely with practitioners to validate our method in a real-life clinical context. We would work on improving the accuracy and interpretability of our model to make it beneficial for both patients and practitioners in real-life health-care process.

A computational aspect to be considered is to investigate in depth on the uncertainty estimation of our model. 
The Bayesian framework offers a way to compute uncertainties when predicting outcome scores, however, in many tasks, these models have shown to be over confident \citep{nalisnick2018deep}. 
It would be useful for to know when a new patient arrives,  how well - with which certainty - we can predict their outcome scores.

The proposed method is a general framework that can be applied to numerous health-care applications involving a long-term healing process after the treatment. In the future, we would collaborate with more health-care departments, test and improve our method in these applications.

\bibliography{sample}

\begin{thebibliography}{38}
\providecommand{\natexlab}[1]{#1}
\providecommand{\url}[1]{\texttt{#1}}
\expandafter\ifx\csname urlstyle\endcsname\relax
  \providecommand{\doi}[1]{doi: #1}\else
  \providecommand{\doi}{doi: \begingroup \urlstyle{rm}\Url}\fi

\bibitem[Alim et~al.(2016)Alim, Svedman, Edman, and Ackermann]{Ackermann16}
Md~Abdul Alim, Simon Svedman, Gunnar Edman, and Paul Ackermann.
\newblock Procollagen markers in microdialysate can predict patient outcome
  after achilles tendon rupture.
\newblock 2016.

\bibitem[Arverud et~al.(2016)Arverud, Anundsson, Hardell, Barreng, Edman,
  Latifi, Labruto, and Ackermann]{arverud2016ageing}
E~Domeij Arverud, Per Anundsson, Eva Hardell, Gunilla Barreng, Gunnar Edman,
  Ali Latifi, Fausto Labruto, and PW~Ackermann.
\newblock Ageing, deep vein thrombosis and male gender predict poor outcome
  after acute achilles tendon rupture.
\newblock \emph{The bone \& joint journal}, 98\penalty0 (12):\penalty0
  1635--1641, 2016.

\bibitem[Blei et~al.(2017)Blei, Kucukelbir, and McAuliffe]{blei2017variational}
David~M Blei, Alp Kucukelbir, and Jon~D McAuliffe.
\newblock Variational inference: A review for statisticians.
\newblock \emph{Journal of the American Statistical Association}, 2017.

\bibitem[Bostick et~al.(2010)Bostick, Jomha, Suchak, and
  Beaupr{\'e}]{bostick2010factors}
Geoff~P Bostick, Nadr~M Jomha, Amar~A Suchak, and Lauren~A Beaupr{\'e}.
\newblock Factors associated with calf muscle endurance recovery 1 year after
  achilles tendon rupture repair.
\newblock \emph{journal of orthopaedic \& sports physical therapy}, 40\penalty0
  (6):\penalty0 345--351, 2010.

\bibitem[Buuren and Groothuis-Oudshoorn(2010)]{buuren2010mice}
S~van Buuren and Karin Groothuis-Oudshoorn.
\newblock mice: Multivariate imputation by chained equations in r.
\newblock \emph{Journal of statistical software}, pages 1--68, 2010.

\bibitem[Cai et~al.(2010)Cai, Cand{\`e}s, and Shen]{cai2010singular}
Jian-Feng Cai, Emmanuel~J Cand{\`e}s, and Zuowei Shen.
\newblock A singular value thresholding algorithm for matrix completion.
\newblock \emph{SIAM Journal on Optimization}, 2010.

\bibitem[Chalapathy et~al.(2016)Chalapathy, Borzeshi, and
  Piccardi]{chalapathy2016bidirectional}
Raghavendra Chalapathy, Ehsan~Zare Borzeshi, and Massimo Piccardi.
\newblock Bidirectional lstm-crf for clinical concept extraction.
\newblock \emph{arXiv preprint arXiv:1611.08373}, 2016.

\bibitem[Domeij-Arverud et~al.(2016)Domeij-Arverud, Anundsson, Hardell,
  Barreng, Edman, Latifi, Labruto, and Ackermann]{domeij2016ageing}
E~Domeij-Arverud, P~Anundsson, E~Hardell, G~Barreng, G~Edman, A~Latifi,
  F~Labruto, and PW~Ackermann.
\newblock Ageing, deep vein thrombosis and male gender predict poor outcome
  after acute achilles tendon rupture.
\newblock \emph{Bone Joint J}, 2016.

\bibitem[Glorot and Bengio(2010)]{glorot2010understanding}
Xavier Glorot and Yoshua Bengio.
\newblock Understanding the difficulty of training deep feedforward neural
  networks.
\newblock In \emph{Proceedings of the thirteenth international conference on
  artificial intelligence and statistics}, 2010.

\bibitem[Horstmann et~al.(2012)Horstmann, Lukas, Merk, Brauner, and
  Mündermann]{deficits10yrhorstmann}
Thomas Horstmann, C~Lukas, J~Merk, Torsten Brauner, and Annegret Mündermann.
\newblock Deficits 10-years after achilles tendon repair.
\newblock 2012.

\bibitem[Huttunen et~al.(2014)Huttunen, Kannus, Rolf, Fell{\"a}nder-Tsai, and
  Mattila]{Huttunen2014AcuteAT}
Tuomas~T. Huttunen, Pekka~A Kannus, Christer~Gustav Rolf,
  Li~Fell{\"a}nder-Tsai, and Ville~M. Mattila.
\newblock Acute achilles tendon ruptures: incidence of injury and surgery in
  sweden between 2001 and 2012.
\newblock \emph{The American journal of sports medicine}, 2014.

\bibitem[Jain et~al.(2010)Jain, Meka, and Dhillon]{jain2010guaranteed}
Prateek Jain, Raghu Meka, and Inderjit~S Dhillon.
\newblock Guaranteed rank minimization via singular value projection.
\newblock In \emph{Advances in Neural Information Processing Systems}, 2010.

\bibitem[Jo et~al.(2017)Jo, Lee, and Palaskar]{jo2017combining}
Yohan Jo, Lisa Lee, and Shruti Palaskar.
\newblock Combining lstm and latent topic modeling for mortality prediction.
\newblock \emph{arXiv preprint arXiv:1709.02842}, 2017.

\bibitem[Keshavan and Oh(2009)]{keshavan2009gradient}
Raghunandan~H Keshavan and Sewoong Oh.
\newblock A gradient descent algorithm on the grassman manifold for matrix
  completion.
\newblock \emph{arXiv preprint arXiv:0910.5260}, 2009.

\bibitem[Keshavan et~al.(2010)Keshavan, Montanari, and Oh]{keshavan2010matrix}
Raghunandan~H Keshavan, Andrea Montanari, and Sewoong Oh.
\newblock Matrix completion from noisy entries.
\newblock \emph{Journal of Machine Learning Research}, 2010.

\bibitem[Lasko(2014)]{lasko2014efficient}
Thomas~A Lasko.
\newblock Efficient inference of gaussian-process-modulated renewal processes
  with application to medical event data.
\newblock In \emph{Uncertainty in artificial intelligence: proceedings of
  the... conference. Conference on Uncertainty in Artificial Intelligence},
  2014.

\bibitem[Lin et~al.(2010)Lin, Chen, and Ma]{lin2010augmented}
Zhouchen Lin, Minming Chen, and Yi~Ma.
\newblock The augmented lagrange multiplier method for exact recovery of
  corrupted low-rank matrices.
\newblock \emph{arXiv preprint arXiv:1009.5055}, 2010.

\bibitem[Ma et~al.(2018)Ma, Tschiatschek, Palla, Lobato, Nowozin, and
  Zhang]{ma2018eddi}
Chao Ma, Sebastian Tschiatschek, Konstantina Palla, Jose Miguel~Hernandez
  Lobato, Sebastian Nowozin, and Cheng Zhang.
\newblock Eddi: Efficient dynamic discovery of high-value information with
  partial vae.
\newblock \emph{arXiv preprint arXiv:1809.11142}, 2018.

\bibitem[Ma et~al.(2011)Ma, Goldfarb, and Chen]{ma2011fixed}
Shiqian Ma, Donald Goldfarb, and Lifeng Chen.
\newblock Fixed point and bregman iterative methods for matrix rank
  minimization.
\newblock \emph{Mathematical Programming}, 2011.

\bibitem[Mazumder et~al.(2010)Mazumder, Hastie, and
  Tibshirani]{mazumder2010spectral}
Rahul Mazumder, Trevor Hastie, and Robert Tibshirani.
\newblock Spectral regularization algorithms for learning large incomplete
  matrices.
\newblock \emph{Journal of machine learning research}, 2010.

\bibitem[Mnih and Salakhutdinov(2008)]{mnih2008probabilistic}
Andriy Mnih and Ruslan~R Salakhutdinov.
\newblock Probabilistic matrix factorization.
\newblock In \emph{Advances in neural information processing systems}, 2008.

\bibitem[Nalisnick et~al.(2018)Nalisnick, Matsukawa, Teh, Gorur, and
  Lakshminarayanan]{nalisnick2018deep}
Eric Nalisnick, Akihiro Matsukawa, Yee~Whye Teh, Dilan Gorur, and Balaji
  Lakshminarayanan.
\newblock Do deep generative models know what they don't know?
\newblock \emph{arXiv preprint arXiv:1810.09136}, 2018.

\bibitem[Neal(2012)]{neal2012bayesian}
Radford~M Neal.
\newblock \emph{Bayesian learning for neural networks}.
\newblock 2012.

\bibitem[Ocepek et~al.(2015)Ocepek, Rugelj, and
  Bosni{\'c}]{ocepek2015improving}
Uro{\v{s}} Ocepek, Jo{\v{z}}e Rugelj, and Zoran Bosni{\'c}.
\newblock Improving matrix factorization recommendations for examples in cold
  start.
\newblock \emph{Expert Systems with Applications}, 2015.

\bibitem[Olsson et~al.(2014)Olsson, Karlsson, Eriksson, Brorsson, Lundberg, and
  Silbernagel]{olsson2014ability}
N~Olsson, J~Karlsson, BI~Eriksson, A~Brorsson, M~Lundberg, and KG~Silbernagel.
\newblock Ability to perform a single heel-rise is significantly related to
  patient-reported outcome after achilles tendon rupture.
\newblock \emph{Scandinavian journal of medicine \& science in sports}, 2014.

\bibitem[Popkes et~al.(2019)Popkes, Overweg, Ercole, Li, Hern{\'a}ndez-Lobato,
  Zaykov, and Zhang]{popkes2019interpretable}
Anna-Lena Popkes, Hiske Overweg, Ari Ercole, Yingzhen Li, Jos{\'e}~Miguel
  Hern{\'a}ndez-Lobato, Yordan Zaykov, and Cheng Zhang.
\newblock Interpretable outcome prediction with sparse bayesian neural networks
  in intensive care.
\newblock \emph{arXiv preprint arXiv:1905.02599}, 2019.

\bibitem[Praxitelous et~al.(2017)Praxitelous, Edman, and
  Ackermann]{praxitelous2017microcirculation}
Praxitelis Praxitelous, Gunnar Edman, and Paul~W Ackermann.
\newblock Microcirculation after achilles tendon rupture correlates with
  functional and patient-reported outcome.
\newblock \emph{Scandinavian journal of medicine \& science in sports}, 2017.

\bibitem[Purushotham et~al.(2017)Purushotham, Meng, Che, and
  Liu]{purushotham2017benchmark}
Sanjay Purushotham, Chuizheng Meng, Zhengping Che, and Yan Liu.
\newblock Benchmark of deep learning models on large healthcare mimic datasets.
\newblock \emph{arXiv preprint arXiv:1710.08531}, 2017.

\bibitem[Scheffer(2002)]{scheffer2002dealing}
Judi Scheffer.
\newblock Dealing with missing data.
\newblock 2002.

\bibitem[Schulam and Saria(2015)]{framework-indiv}
Peter Schulam and Suchi Saria.
\newblock A framework for individualizing predictions of disease trajectories
  by exploiting multi-resolution structure.
\newblock In \emph{Proceedings of the 28th International Conference on Neural
  Information Processing Systems}, 2015.

\bibitem[Shi et~al.(2016)Shi, Zhu, Philip, Huang, Wang, Mao, and
  Chen]{shi2016temporal}
Weiwei Shi, Yongxin Zhu, S~Yu Philip, Tian Huang, Chang Wang, Yishu Mao, and
  Yufeng Chen.
\newblock Temporal dynamic matrix factorization for missing data prediction in
  large scale coevolving time series.
\newblock \emph{IEEE Access}, 2016.

\bibitem[Stern et~al.(2009)Stern, Herbrich, and Graepel]{matchbox}
David Stern, Ralf Herbrich, and Thore Graepel.
\newblock Matchbox: Large scale bayesian recommendations.
\newblock In \emph{Proceedings of the 18th International World Wide Web
  Conference}, 2009.

\bibitem[Suresh et~al.(2017)Suresh, Hunt, Johnson, Celi, Szolovits, and
  Ghassemi]{suresh2017clinical}
Harini Suresh, Nathan Hunt, Alistair Johnson, Leo~Anthony Celi, Peter
  Szolovits, and Marzyeh Ghassemi.
\newblock Clinical intervention prediction and understanding with deep neural
  networks.
\newblock In \emph{Machine Learning for Healthcare Conference}, pages 322--337,
  2017.

\bibitem[Tran et~al.(2016)Tran, Kucukelbir, Dieng, Rudolph, Liang, and
  Blei]{tran2016edward}
Dustin Tran, Alp Kucukelbir, Adji~B. Dieng, Maja Rudolph, Dawen Liang, and
  David~M. Blei.
\newblock Edward: A library for probabilistic modeling, inference, and
  criticism, 2016.

\bibitem[Troyanskaya et~al.(2001)Troyanskaya, Cantor, Sherlock, Brown, Hastie,
  Tibshirani, Botstein, and Altman]{troyanskaya2001missing}
Olga Troyanskaya, Michael Cantor, Gavin Sherlock, Pat Brown, Trevor Hastie,
  Robert Tibshirani, David Botstein, and Russ~B Altman.
\newblock Missing value estimation methods for dna microarrays.
\newblock \emph{Bioinformatics}, 2001.

\bibitem[Valkering et~al.(2017)Valkering, Aufwerber, Ranuccio, Lunini, Edman,
  and Ackermann]{valkering2017functional}
Kars~P Valkering, Susanna Aufwerber, Francesco Ranuccio, Enricomaria Lunini,
  Gunnar Edman, and Paul~W Ackermann.
\newblock Functional weight-bearing mobilization after achilles tendon rupture
  enhances early healing response: a single-blinded randomized controlled
  trial.
\newblock \emph{Knee Surgery, Sports Traumatology, Arthroscopy}, 2017.

\bibitem[Zhang et~al.(2015)Zhang, Gartrell, Minka, Zaykov, and
  Guiver]{groupbox}
Cheng Zhang, Mike Gartrell, Thomas Minka, Yordan Zaykov, and John Guiver.
\newblock Groupbox: A generative model for group recommendation.
\newblock Technical Report MSR-TR-2015-61, 2015.

\bibitem[Zhang et~al.(2017)Zhang, Butepage, Kjellstrom, and
  Mandt]{zhang2017advances}
Cheng Zhang, Judith Butepage, Hedvig Kjellstrom, and Stephan Mandt.
\newblock Advances in variational inference.
\newblock \emph{arXiv preprint arXiv:1711.05597}, 2017.

\end{thebibliography}

\newpage
\appendix
\section{Variables}

\subsection{Predictors}

\begin{longtable}[l]{| l | l |}
\hline
0 & ID \\ \hline 1 & Study \\ \hline 2 & Gender \\ \hline 3 & Age \\ \hline 4 & DIC\_age\_40 \\ \hline 5 & Length \\ \hline 6 & Weight \\ \hline 7 & BMI \\ \hline 8 & DIC\_BMI\_27 \\ \hline 9 & Smoker \\ \hline 10 & ln\_Age \\ \hline 11 & ln\_Lenght \\ \hline 12 & ln\_Weight \\ \hline 13 & ln\_BMI \\ \hline 14 & Inj\_side \\ \hline 15 & Complication \\ \hline 16 & Paratenon \\ \hline 17 & Fascia \\ \hline 18 & PDS \\ \hline 19 & Surg\_comp \\ \hline 20 & Treatment\_Group \\ \hline 21 & Ort\_B \\ \hline 22 & Plast \\ \hline 23 & Healthy\_control \\ \hline 24 & Plast\_foot \\ \hline 25 & Vacoped \\ \hline 26 & VTIS \\ \hline 27 & TTS \\ \hline 28 & ln\_TTS \\ \hline 29 & TTS\_no\_of\_24h\_cycles \\ \hline 30 & ln\_TTS\_no\_of\_24h\_cycles \\ \hline 31 & TEST\_TTS\_48h\_POL \\ \hline 32 & TEST\_TTS\_24h\_POL \\ \hline 33 & TEST\_TTS\_12h\_POL \\ \hline 34 & DIC\_TTS\_by\_VTIS\_median \\ \hline 35 & TRICH\_48\_84\_TTS \\ \hline 36 & TRICH\_48\_96\_TTS \\ \hline 37 & TRICH\_48\_72\_TTS \\ \hline 38 & NEW\_Time\_er\_to\_op\_start \\ \hline 39 & DIC\_NEW\_Time\_er\_to\_op\_start \\ \hline 40 & DIC3\_NEW\_Time\_er\_to\_op\_start \\ \hline 41 & OLD\_Time\_er\_to\_op\_start \\ \hline 42 & Op\_time \\ \hline 43 & DIC\_op\_34min \\ \hline 44 & Op\_B\_Dic \\ \hline 45 & OP\_GBG\_dic \\ \hline 46 & Op\_time\_dic \\ \hline 47 & OP\_NR \\ \hline 48 & EXP \\ \hline 49 & Q\_RANK\_B \\ \hline 50 & Q\_RANK\_ABCD \\ \hline 51 & NR\_of\_Op \\ \hline 52 & DIC\_nr\_OP \\ \hline 53 & ASS\_Y\_N \\ \hline 54 & DIK\_SPEC \\ \hline 55 & PP\_IPC\_Study\_B \\ \hline 56 & Pump\_pat\_reg \\ \hline 57 & Pump\_reg \\ \hline 58 & Highest\_pump\_reg \\ \hline 59 & DIC\_86h\_highest\_pump\_reg \\ \hline 60 & Pump\_comp \\ \hline 61 & Incl\_Excl \\ \hline 62 & DVT\_2 \\ \hline 63 & DVT\_6w \\ \hline 64 & DVT\_2w\_and\_6w \\ \hline 65 & DVT\_2w\_or\_6w \\ \hline 66 & DVT\_8w \\ \hline 67 & Any\_dvt \\ \hline 68 & Inf\_2w \\ \hline 69 & Inf\_6w \\ \hline 70 & Any\_inf \\ \hline 71 & Rerupture \\ \hline 72 & Adeverse\_events\_1 \\ \hline 73 & Adeverse\_events\_2 \\ \hline 74 & Adeverse\_events\_3 \\ \hline 75 & Adeverse\_events\_4 \\ \hline 76 & Preinjury \\ \hline 77 & Post\_op \\ \hline 78 & D\_PAS \\ \hline 79 & Preinj\_2cl \\ \hline 80 & Preinj\_3cl \\ \hline 81 & Post\_op\_2cl \\ \hline 82 & Con\_Power\_I \\ \hline 83 & Con\_Power\_U \\ \hline 84 & LSI\_Con\_Power \\ \hline 85 & Total\_work\_I \\ \hline 86 & Total\_work\_U \\ \hline 87 & NEW\_LSI\_Total\_work \\ \hline 88 & LSI\_Total\_work \\ \hline 89 & Repetition\_I \\ \hline 90 & Repetition\_U \\ \hline 91 & LSI\_Repetitions \\ \hline 92 & Height\_Max\_I \\ \hline 93 & Height\_Max\_U \\ \hline 94 & LSI\_Height \\ \hline 95 & Height\_A\_I \\ \hline 96 & Height\_A\_U \\ \hline 97 & LSI\_Height\_Ave \\ \hline 98 & Ecc\_Power\_I \\ \hline 99 & Height\_Min\_I \\ \hline 100 & Ecc\_Power\_U \\ \hline 101 & Height\_Min\_U \\ \hline 102 & LSI\_Height\_2cl \\ \hline 103 & LSI\_Height\_Min \\ \hline 104 & LSI\_Ecc\_Power \\ \hline 105 & Muscle\_vein\_thrombosis\_2 \\ \hline 106 & Thompson\_2 \\ \hline 107 & Wound\_2 \\ \hline 108 & Podometer\_day1 \\ \hline 109 & Podometer\_day2 \\ \hline 110 & Podometer\_day3 \\ \hline 111 & Podometer\_day4 \\ \hline 112 & Podometer\_day5 \\ \hline 113 & Podometer\_day6 \\ \hline 114 & Podometer\_day7 \\ \hline 115 & Podometer\_day8 \\ \hline 116 & Podometer\_day9 \\ \hline 117 & Podometer\_day10 \\ \hline 118 & Podometer\_day11 \\ \hline 119 & Podometer\_day12 \\ \hline 120 & Podometer\_day13 \\ \hline 121 & Podometer\_day14 \\ \hline 122 & Podometer\_day15 \\ \hline 123 & Podometer\_day16 \\ \hline 124 & Mean\_podometer \\ \hline 125 & Total\_podometer \\ \hline 126 & DIC\_podometer\_16500 \\ \hline 127 & Podometer\_on\_day\_of\_microdialysis \\ \hline 128 & Podometer\_on\_day\_minus\_1 \\ \hline 129 & Podometer\_on\_day\_minus\_2 \\ \hline 130 & Subjective\_load\_day1 \\ \hline 131 & Subjective\_load\_day2 \\ \hline 132 & Subjective\_load\_day3 \\ \hline 133 & Subjective\_load\_day4 \\ \hline 134 & Subjective\_load\_day5 \\ \hline 135 & Subjective\_load\_day6 \\ \hline 136 & Subjective\_load\_day7 \\ \hline 137 & Subjective\_load\_day8 \\ \hline 138 & Subjective\_load\_day9 \\ \hline 139 & Subjective\_load\_day10 \\ \hline 140 & Subjective\_load\_day11 \\ \hline 141 & Subjective\_load\_day12 \\ \hline 142 & Subjective\_load\_day13 \\ \hline 143 & Subjective\_load\_day14 \\ \hline 144 & Subjective\_load\_day15 \\ \hline 145 & Subjective\_load\_day16 \\ \hline 146 & Mean\_subjective\_load \\ \hline 147 & DIC\_43\_Mean\_subjective\_load \\ \hline 148 & Days\_until\_microdialysis \\ \hline 149 & Load\_on\_day\_of\_microdialysis \\ \hline 150 & Load\_on\_day\_minus\_1 \\ \hline 151 & Load\_on\_day\_minus\_2 \\ \hline 152 & Number\_of\_days\_with\_load\_prior\_to\_microdialysis \\ \hline 153 & DIC\_13\_days\_with\_load \\ \hline 154 & VAS\_day1\_act \\ \hline 155 & VAS\_day1\_pas \\ \hline 156 & VAS\_day2\_act \\ \hline 157 & VAS\_day2\_pas \\ \hline 158 & VAS\_day3\_act \\ \hline 159 & VAS\_day3\_pas \\ \hline 160 & VAS\_day4\_act \\ \hline 161 & VAS\_day4\_pas \\ \hline 162 & VAS\_day5\_act \\ \hline 163 & VAS\_day5\_pas \\ \hline 164 & VAS\_day6\_act \\ \hline 165 & VAS\_day6\_pas \\ \hline 166 & VAS\_day7\_act \\ \hline 167 & VAS\_day7\_pas \\ \hline 168 & VAS\_injured\_2weeks \\ \hline 169 & VAS\_control\_2weeks \\ \hline 170 & Calf\_circumference\_injured\_1 \\ \hline 171 & Calf\_circumference\_injured\_mean \\ \hline 172 & Calf\_circumference\_control\_1 \\ \hline 173 & Calf\_circumference\_control\_mean \\ \hline 174 & Plantar\_flexion\_injured\_1 \\ \hline 175 & Plantar\_flexion\_injured\_2 \\ \hline 176 & Plantar\_flexion\_injured\_3 \\ \hline 177 & Plantar\_flexion\_injured\_mean \\ \hline 178 & Plantar\_flexion\_control\_1 \\ \hline 179 & Plantar\_flexion\_control\_2 \\ \hline 180 & Plantar\_flexion\_control\_3 \\ \hline 181 & Plantar\_flexion\_control\_mean \\ \hline 182 & Dorsal\_flexion\_injured\_1 \\ \hline 183 & Dorsal\_flexion\_injured\_2 \\ \hline 184 & Dorsal\_flexion\_injured\_3 \\ \hline 185 & Dorsal\_flexion\_injured\_mean \\ \hline 186 & Dorsal\_flexion\_control\_1 \\ \hline 187 & Dorsal\_flexion\_control\_2 \\ \hline 188 & Dorsal\_flexion\_control\_3 \\ \hline 189 & Dorsal\_flexion\_control\_mean \\ \hline 190 & Q1 \\ \hline 191 & Q2 \\ \hline 192 & Q3 \\ \hline 193 & Q4 \\ \hline 194 & Q5 \\ \hline 195 & EQ5D\_ix \\ \hline 196 & VAS \\ \hline 197 & VAS\_2 \\ \hline 198 & Gluc2\_2\_i \\ \hline 199 & Gluc2\_3\_i \\ \hline 200 & Gluc2\_4\_i \\ \hline 201 & GLUC\_injured\_mean \\ \hline 202 & Gluc2\_2\_c \\ \hline 203 & Gluc2\_3\_c \\ \hline 204 & Gluc2\_4\_c \\ \hline 205 & GLUC\_control\_mean \\ \hline 206 & Lact2\_2\_i \\ \hline 207 & Lact2\_3\_i \\ \hline 208 & Lact2\_4\_i \\ \hline 209 & LAC\_injured\_mean \\ \hline 210 & Lact2\_2\_c \\ \hline 211 & Lact2\_3\_c \\ \hline 212 & Lact2\_4\_c \\ \hline 213 & LAC\_control\_mean \\ \hline 214 & Pyr2\_2\_i \\ \hline 215 & Pyr2\_3\_i \\ \hline 216 & Pyr2\_4\_i \\ \hline 217 & PYR\_injured\_mean \\ \hline 218 & Pyr2\_2\_c \\ \hline 219 & Pyr2\_3\_c \\ \hline 220 & Pyr2\_4\_c \\ \hline 221 & PYR\_control\_mean \\ \hline 222 & Glyc2\_2\_i \\ \hline 223 & Glyc2\_3\_i \\ \hline 224 & Glyc2\_4\_i \\ \hline 225 & GLY\_injured\_mean \\ \hline 226 & Glyc2\_2\_c \\ \hline 227 & Glyc2\_3\_c \\ \hline 228 & Glyc2\_4\_c \\ \hline 229 & GLY\_control\_mean \\ \hline 230 & Glut2\_2\_i \\ \hline 231 & Glut2\_3\_i \\ \hline 232 & Glut2\_4\_i \\ \hline 233 & GLUT\_injured\_mean \\ \hline 234 & Glut2\_2\_c \\ \hline 235 & Glut2\_3\_c \\ \hline 236 & Glut2\_4\_c \\ \hline 237 & GLUT\_control\_mean \\ \hline 238 & Lac2\_Pyr2\_ratio\_2\_i \\ \hline 239 & Lac2\_Pyr2\_ratio\_3\_i \\ \hline 240 & Lac2\_Pyr2\_ratio\_4\_i \\ \hline 241 & LAC2\_PYR2\_ratio\_injured\_mean \\ \hline 242 & Lac2\_Pyr2\_ratio\_2\_c \\ \hline 243 & Lac2\_Pyr2\_ratio\_3\_c \\ \hline 244 & Lac2\_Pyr2\_ratio\_4\_c \\ \hline 245 & LAC2\_PYR2\_ratio\_control\_mean \\ \hline 246 & PINP\_injured \\ \hline 247 & PIIINP\_injured \\ \hline 248 & Bradford\_injured \\ \hline 249 & PINP\_normalized\_Injured \\ \hline 250 & PIIINP\_normalized\_injured \\ \hline 251 & PINP\_uninjured \\ \hline 252 & PIIINP\_uninjured \\ \hline 253 & Bradford\_uninjured \\ \hline 254 & PINP\_normalized\_Uninjured \\ \hline 255 & PIIINP\_normalized\_uninjured \\ \hline 256 & Collagen \\ \hline 257 & Glut\_2\_inj\_values \\ \hline 258 & P\_ratio\_inj \\ \hline 259 & DIC\_PIIINP \\ \hline 260 & DIC\_PIIINP\_3 \\ \hline 261 & P\_ratio\_uninj \\ \hline 262 & FIL\_OP\_STUDY \\ \hline 263 & Gly\_inv \\ \hline 264 & RF\_injured \\ \hline 265 & RF\_uninjured \\ \hline 266 & BZ\_injured \\ \hline 267 & BZ\_uninjured \\ \hline 268 & MF\_injured \\ \hline 269 & MF\_uninjured \\ \hline 270 & T\_RF\_injured \\ \hline 271 & T\_RF\_uninjured \\ \hline 272 & T\_MF\_injured \\ \hline 273 & T\_MF\_uninjured \\ \hline 274 & T\_HR\_injured \\ \hline 275 & T\_HR\_uninjured \\ \hline 276 & Ratio\_MF\_RF\_injured \\ \hline 277 & Ratio\_MF\_RF\_uninjured \\ \hline 278 & B1\_D66 \\ \hline 279 & Sthlm\_gbg \\ \hline 280 & stepsxload\_day1 \\ \hline 281 & stepsxload\_day2 \\ \hline 282 & stepsxload\_day3 \\ \hline 283 & stepsxload\_day4 \\ \hline 284 & stepsxload\_day5 \\ \hline 285 & stepsxload\_day6 \\ \hline 286 & stepsxload\_day7 \\ \hline 287 & stepsxload\_day8 \\ \hline 288 & stepsxload\_day9 \\ \hline 289 & stepsxload\_day10 \\ \hline 290 & stepsxload\_day11 \\ \hline 291 & stepsxload\_day12 \\ \hline 292 & stepsxload\_day13 \\ \hline 293 & stepsxload\_day14 \\ \hline 294 & stepsxload\_day15 \\ \hline 295 & stepsxload\_day16 \\ \hline 296 & INC\_A42 \\ \hline 
\end{longtable}

\subsection{Scores}

\begin{longtable}[l]{| l | l |}
\hline 
0 & DIC\_TTS\_by\_valid\_ATRS80\_median \\ \hline 1 & Control\_1yr \\ \hline 2 & Heel\_rise\_average\_height\_injured\_6mo \\ \hline 3 & Heel\_rise\_average\_height\_control\_6mo \\ \hline 4 & difference\_heel\_raise\_6mo \\ \hline 5 & Heel\_rise\_average\_height\_injured\_1yr \\ \hline 6 & Heel\_rise\_average\_height\_control\_1yr \\ \hline 7 & difference\_heel\_raise\_1yr \\ \hline 8 & ATRS\_3\_Strenght \\ \hline 9 & ATRS\_3\_tired \\ \hline 10 & ATRS\_3\_stiff \\ \hline 11 & ATRS\_3\_pain \\ \hline 12 & ATRS\_3\_ADL \\ \hline 13 & ATRS\_3\_Surface \\ \hline 14 & ATRS\_3\_stairs \\ \hline 15 & ATRS\_3\_run \\ \hline 16 & ATRS\_3\_jump \\ \hline 17 & ATRS\_3\_phys \\ \hline 18 & ATRS\_3\_Sum \\ \hline 19 & ATRS\_item1\_6month \\ \hline 20 & ATRS\_item2\_6month \\ \hline 21 & ATRS\_item3\_6month \\ \hline 22 & ATRS\_item4\_6month \\ \hline 23 & ATRS\_item5\_6month \\ \hline 24 & ATRS\_item6\_6month \\ \hline 25 & ATRS\_item7\_6month \\ \hline 26 & ATRS\_item8\_6month \\ \hline 27 & ATRS\_item9\_6month \\ \hline 28 & ATRS\_item10\_6month \\ \hline 29 & ATRS\_total\_score\_6month \\ \hline 30 & ATRS\_12\_strength \\ \hline 31 & ATRS\_12\_tired \\ \hline 32 & ATRS\_12\_stiff \\ \hline 33 & ATRS\_12\_pain \\ \hline 34 & ATRS\_12\_ADL \\ \hline 35 & ATRS\_12\_Surface \\ \hline 36 & ATRS\_12\_stairs \\ \hline 37 & ATRS\_12\_run \\ \hline 38 & ATRS\_12\_jump \\ \hline 39 & ATRS\_12\_phys \\ \hline 40 & ATRS\_12m \\ \hline 41 & valid\_ATRS\_12m \\ \hline 42 & ATRS\_2cl \\ \hline 43 & FAOS\_3\_Pain \\ \hline 44 & FAOS\_3\_Symptom \\ \hline 45 & FAOS\_3\_ADL \\ \hline 46 & FAOS\_3\_sport\_rec \\ \hline 47 & FAOS\_3\_QOL \\ \hline 48 & FAOS\_6\_Symptom \\ \hline 49 & FAOS\_6\_Pain \\ \hline 50 & FAOS\_6\_ADL \\ \hline 51 & FAOS\_6\_Sport\_Rec \\ \hline 52 & FAOS\_6\_QOL \\ \hline 53 & FAOS\_12\_Pain \\ \hline 54 & FAOS\_12\_Symptom \\ \hline 55 & FAOS\_12\_ADL \\ \hline 56 & FAOS\_12\_Sport\_Rec \\ \hline 57 & FAOS\_12\_QOL \\ \hline 58 & ATRS\_12\_pain\_log \\ \hline 59 & ATRS\_12\_ADL\_log \\ \hline 60 & ATRS\_12\_surface\_log \\ \hline 61 & ATRS\_12\_phys\_log \\ \hline 62 & FAOS\_12\_pain\_log \\ \hline 
\end{longtable}
\end{document}